\documentclass[12pt]{article}  
\usepackage{epsfig}

\newcommand{\tr}{{\rm Tr}}

\def\beq{\begin{equation}}
\def\eeq{\end{equation}}
\def\to{\rightarrow}

\def\bsg{\ifmmode B\to X_s\gamma\else $B\to X_s\gamma$\fi}
\def\bsll{\ifmmode B\to X_s\ell^+\ell^-\else $B\to X_s\ell^+\ell^-$\fi}
\def\bstt{\ifmmode B\to X_s\tau^+\tau^-\else $B\to X_s\tau^+\tau^-$\fi}
\def\shat{\ifmmode \hat{s}\else $\hat{s}$\fi}

\newcommand{\newc}{\newcommand}

\newc{\lcal}{\int {\cal L}dt}
 
\newc{\LSP}{{\chi^0_1}}
\newc{\stauR}{{\tilde \tau_R}}
\newc{\stau}{{\tilde \tau_1}}
\newc{\mstop}{m_{\tilde{t}}}
\newc{\mHpm}{m_{H^\pm}}
\newc{\gsim}{\lower.7ex\hbox{$\;\stackrel{\textstyle>}{\sim}\;$}}
\newc{\lsim}{\lower.7ex\hbox{$\;\stackrel{\textstyle<}{\sim}\;$}}
\newc{\ie}{{\it i.e.}}          
\newc{\etal}{{\it et al.}}
\newc{\eg}{{\it e.g.}}          
\newc{\kev}{\hbox{\rm\,keV}}            
\newc{\mev}{\hbox{\rm\,MeV}}            
\newc{\gev}{\hbox{\rm\,GeV}}            
\newc{\tev}{\hbox{\rm\,TeV}}
\newc{\xpb}{\hbox{\rm\, pb}}
\newc{\xfb}{\hbox{\rm\, fb}}

\newc{\mtop}{m_t}
\newc{\mbot}{m_b}
\newc{\mz}{M_Z}
\newc{\mw}{M_W}
\newc{\alphasmz}{\alpha_s(M_Z)}
\newc{\swsq}{\sin^2\theta_W}
\newc{\cwsq}{\cos^2\theta_W}
\newc{\tw}{\tan\theta_W}
\newc{\cw}{\cos\theta_W}
\newc{\sw}{\sin\theta_W}
\newc{\BR}{\hbox{\rm BR}}
\newc{\zbb}{Z\to b\bar}
\newc{\Gb}{\Gamma (Z\to b\bar b)}
\newc{\Gh}{\Gamma (Z\to \hbox{\rm hadrons})}
\newc{\rbsm}{R_b^\hbox{\rm sm}}
\newc{\rbsusy}{R_b^\hbox{\rm susy}}
\newc{\drb}{\delta R_b}

\newc{\sgn}{\mbox{sgn}}

\def\eq#1{eq.~(\ref{#1})}
\def\fig#1{fig.~\ref{#1}}
\def\beqa{\begin{eqnarray}}
\def\eeqa{\end{eqnarray}}

\def\mgut{M_{\rm GUT}}
\def\agut{\alpha_{\rm GUT}}
\def\gtilu{{\tilde g}_u}
\def\gtild{{\tilde g}_d}
\def\gtilup{{\tilde g}_u^\prime}
\def\gtildp{{\tilde g}_d^\prime}
\def\gtiluq{{\tilde g}_u^2}
\def\gtildq{{\tilde g}_d^2}
\def\gtilupq{{\tilde g}_u^{\prime 2}}
\def\gtildpq{{\tilde g}_d^{\prime 2}}
\def\gtiluc{{\tilde g}_u^3}
\def\gtildc{{\tilde g}_d^3}
\def\gtilupc{{\tilde g}_u^{\prime 3}}
\def\gtildpc{{\tilde g}_d^{\prime 3}}
\def\gtiluqq{{\tilde g}_u^4}
\def\gtildqq{{\tilde g}_d^4}

\def\beq{\begin{equation}}
\def\eeq{\end{equation}}
\def\bea{\begin{eqnarray}}
\def\eea{\end{eqnarray}}
\def\slashchar#1{\setbox0=\hbox{$#1$}           
   \dimen0=\wd0                                 
   \setbox1=\hbox{/} \dimen1=\wd1               
   \ifdim\dimen0>\dimen1                        
      \rlap{\hbox to \dimen0{\hfil/\hfil}}      
      #1                                        
   \else                                        
      \rlap{\hbox to \dimen1{\hfil$#1$\hfil}}   
      /                                         
   \fi}                                         %
\catcode`@=11
\long\def\@caption#1[#2]#3{\par\addcontentsline{\csname
  ext@#1\endcsname}{#1}{\protect\numberline{\csname
  the#1\endcsname}{\ignorespaces #2}}\begingroup
    \small
    \@parboxrestore
    \@makecaption{\csname fnum@#1\endcsname}{\ignorespaces #3}\par
  \endgroup}
\catcode`@=12

\begin{document}

\baselineskip=18pt

\setcounter{footnote}{0}
\setcounter{figure}{0}
\setcounter{table}{0}

\begin{titlepage}
June 2004 \hspace*{\fill}
CERN-TH/2004-100

\begin{center}
\vspace{1cm}

{\Large \bf 

Split Supersymmetry}

\vspace{0.8cm}

{\bf G. F. Giudice} and {\bf A. Romanino}

\vspace{.5cm}

{\it CERN, Theory Division, CH-1211 Geneva 23, Switzerland}

\end{center}
\vspace{1cm}

\begin{abstract}
\medskip
The naturalness criterion applied to the cosmological constant implies
a new-physics threshold at $10^{-3}$~eV. Either the naturalness criterion
fails, or this threshold does not influence particle
dynamics at higher energies. 
It has been suggested that the Higgs naturalness problem may
follow the same fate. We investigate this possibility and,  
abandoning the hierarchy problem, we use unification and
dark matter as the only guiding principles.
The model recently proposed by
Arkani-Hamed and Dimopoulos emerges as a very interesting option. We
study it in detail, analysing its structure, and the conditions for
obtaining unification and dark matter.

\end{abstract}

\bigskip
\bigskip


\end{titlepage}


\section{Introduction}
\label{sec1}

For decades the naturalness (or hierarchy) problem of the Higgs mass term 
has been
the guiding principle to construct theories beyond the Standard Model (SM).
The criterion of naturalness has the exciting implication that the SM should
stop to be valid at a scale around the TeV, and new dynamics should occur
at energies reachable by present or near-future colliders. Although no clear
indications for any SM failure at electroweak energies has emerged so far, 
a conclusive resolution of this issue has to wait for the LHC.

From a field-theoretical point of 
view, the cosmological constant problem appears 
to be very similar to the naturalness problem of the Higgs mass, since both
of them are related to ultraviolet power divergences. The same naturalness
criterion, applied to the cosmological costant, leads to the existence
of a threshold of new dynamics at $10^{-3}$~eV. We do not know if some
hidden dynamics actually takes place at that scale, or if the resolution
of the problem comes without any modification of the dynamics. What we know
is that present particle physics calculations, valid at energies much
larger than $10^{-3}$~eV, can be safely performed by setting
the cosmological constant to zero and ignore any effect caused by the mechanism
ultimately responsible for the solution of the problem. This fact has
been justified by invoking
the anthropic principle~\cite{wein}, which could be operating in presence 
of a large number of meta-stable vacua, as in string theory~\cite{string}.

It is conceivable to ponder whether such an explanation could also apply to
the hierarchy problem, imagining a 
mechanism (not necessarily based on
the anthropic
principle) which 
allows to extrapolate SM calculations to energies much larger than the TeV,
without the need of
introducing new dynamics, besides the Higgs.

At first sight, this sounds like a devastating proposal. But, if we are
willing to abandon the hierarchy problem, we can try to use other 
clues to drive the search for the theory
beyond the SM. Gauge coupling unification could be one such clue: it is 
motivated by a theory that addresses questions 
related to the fundamental structure of the SM particle content. 

The failure of exact unification of gauge couplings in the SM suggests
the existence of new particles, belonging to incomplete GUT irreps, which
mend the mismatch. It is well known that low-energy supersymmetry 
provides precisely the necessary particles with the appropriate quantum
numbers.
Recently, Arkani-Hamed and Dimopoulos~\cite{savas},
setting aside the hierarchy problem, have noticed that gauge-coupling
unification can be achieved in a supersymmetric model where all scalars, 
but one Higgs doublet, are much
heavier than the electroweak scale. Most of the unpleasent aspects
of supersymmetry (excessive flavour and CP violation, fast 
dimension-5 proton decay, tight constraints on the Higgs mass)
are eliminated, but the unification is retained. 

If supersymmetry plays no r\^ole in solving 
the hierarchy problem, there is no reason
to insist that the spectrum is (partially) supersymmetric. Therefore, in this
paper we perform a general analysis of the particle content that has to
be added to the SM to obtain gauge-coupling unification. We assume that the new
particles have masses around the weak scale, and postulate the absence of
thresholds at intermediate energies or various stages of gauge symmetry
breaking. Why should the new states appear at the weak scale, if the 
hierarchy problem is not the guiding principle of the analysis? 

Here we can use a second observational clue: the evidence for dark matter
and the observation that a particle with weak cross section and mass 
around the Fermi scale is a natural candidate for it. We stress that
the unification and dark-matter arguments are not in general sufficient
to insure that new physics be within the LHC discovery reach, contrary to the
naturalness criterion. Nevertheless, as we will show in our analysis,
in some cases there are interesting experimental consequences to be
investigated. 

In particular, as explained in sect.~\ref{versus}, 
we find that the model proposed in ref.~\cite{savas},
which we call Split Supersymmetry, emerges as 
more justified than other 
non-supersymmetric scenarios. For this reason, in sect.~\ref{sec3}, we
perform a careful analysis of the predictions of Split Supersymmetry,
extending the results of ref.~\cite{savas}. We perform a two-loop analysis
of gauge coupling unification and discuss the $m_b/m_\tau$ relation. 
We discuss the predictions for gaugino masses and couplings and for the
Higgs mass. Next, we link the gaugino and higgsino masses to the
weak scale by computing the relic abundance of the lightest neutralino
and by requiring that it constitutes the dark matter. Finally, we briefly
discuss signals at high-energy colliders.

\section{Conditions for Gauge-Coupling Unification}
\label{sec2}

In this section we want to classify all possible particle contents that
lead to a successful gauge-coupling unification, under the two following
assumptions: {\it (i)} grand unification into a simple group occurs with 
no intermediate stages of symmetry breaking; {\it (ii)} the new particles
have masses comparable with the electroweak scale. We do not necessarily
assume low-energy supersymmetry.

Considering one-loop evolution of gauge couplings, the GUT predictions
of the strong coupling $\alpha_s$, and of the unification scale
$\mgut$ and coupling $\agut$ are \bea \alpha_s^{-1}&=&
\alpha^{-1}\left[ \swsq +\frac{3-8\swsq}{5}
  \left( \frac{b_3-b_2}{b_1-b_2}\right)\right] ,\\
\ln \frac{\mgut}{\mz}&=&\frac{2\pi \left(3-8\swsq \right)}
{5\alpha (b_1-b_2)},\\
\agut &=& \frac{5\alpha (b_1-b_2)}{5\swsq b_1-3\cwsq b_2}, \eea \bea
b_3 &=& \frac{1}{3}\left( 4N_g -33+N_3 \right) ,\\
b_2 &=& \frac{1}{3}\left( 4N_g-22 +\frac{n_H}{2}+N_2 \right) ,\\
b_1 &=& \frac{1}{3}\left( 4N_g +\frac{3n_H}{10}+N_1 \right) .  \eea
Here $N_g$ counts the contribution to the $\beta$-functions from
complete $SU(5)$ irreps, and it is normalized such that the 3 families
of SM quarks and leptons give $N_g=3$. In the case of low-energy
supersymmetry $N_g=9/2$, because of the extra contributions from
squarks and sleptons (which count 1/2 for each generation). However,
notice that $N_g$ does not affect the predictions of $\alpha_s$ and
$\mgut$, since its contribution cancels in the difference of two
$\beta$-function coefficients. Next, $n_H$ counts the number of Higgs
doublets, and we take $n_H=1$. Finally, $N_i$ give the contributions
from matter in incomplete GUT multiplets. The normalization is such
that, for two-component fermions, $N_i$ is the Dynkin index of the
irrep\footnote{In the case of $SU(N)$ groups, the Dynkin index is 1
  for the fundamental, $2N$ for the adjoint, and $N+2$ for the
  symmetric product of two fundamentals. In the case of $U(1)_Y$, we
  define the Dynkin index as $(3/10)N_fY^2$, where $N_f$ is the number
  of multiplet components and $Y$ is the hypercharge ($Y^2=1$ for the
  SM Higgs).} and, for complex scalars, $N_i$ is half the Dynkin
index. With this definition, all $N_i$ are positive, and $N_{2,3}$ are
integers for chiral fermions, even integers for fermions in real
irreps, and half integers for complex scalars.  The same holds for the
quantity $5N_1$, if the representation can be embedded in
$SU(5)$.\footnote{\label{obs} This can be proven by using the
  following observation. Let us consider an irrep of $SU(N)$ in a
  subspace $\Sigma$ of a tensor product of fundamentals. Every linear
  operator $X$ acting on the space of the fundamental representation
  induces a linear operator $X_\Sigma$ in $\Sigma$, defined as the sum
  of the action of $X$ on the single tensor factors. We then have
   \begin{eqnarray}
   \label{eq:obs}
   &\tr X_\Sigma = r_\Sigma \tr X \nonumber \\
   &\tr X_\Sigma Y_\Sigma = p_\Sigma \tr X \tr Y + q_\Sigma \tr X Y \,,
   \end{eqnarray}
   where $p_\Sigma$, $q_\Sigma$, $r_\Sigma$ are non-negative integers
   independent of $X$, $Y$. Let us now consider a SM chiral fermion irrep from
   $SU(5)$. That can be written as the tensor product of an irrep of
   $SU(2)$ with dimension $D_2$ and an irrep of $SU(3)$ with dimension
   $D_3$. Moreover, the hypercharge gives 1 on each $SU(2)$ factor and
   -2/3 on each $SU(3)$ factor. By using eq.~\ref{eq:obs} we then find
   \begin{eqnarray}
     \label{eq:obs2}
     &N_3 = q_3 D_2, \qquad N_2 = q_2 D_3, \nonumber \\
     &\frac{5}{2} N_1 = \frac{3}{4} \tr Y^2 = (3p_3+q_3)D_2
     +(2p_2+q_2)D_3 -2r_2 r_3 \,,
   \end{eqnarray}
   where $p_2,q_2,r_2$, $p_3,q_3,r_3$ are given by eqs.~\ref{eq:obs}.
   Therefore, $N_3$, $N_2$ and $5N_1/2$ are integers.}

\begin{figure}
\begin{center}
\epsfig{file=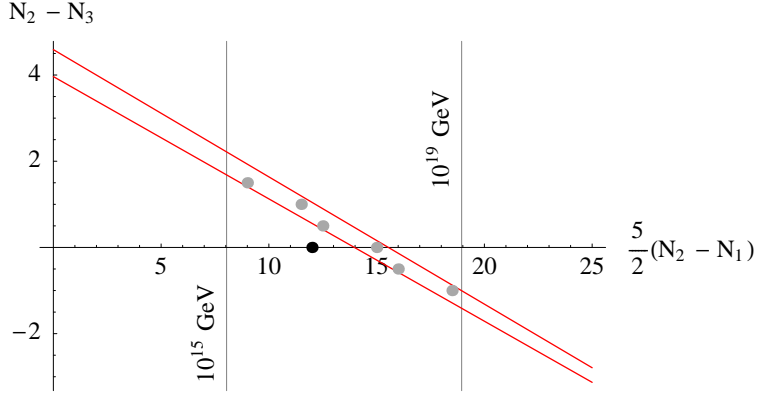,width=0.6\textwidth}
\end{center}
\caption{On the plane of the Dynkin indices of new
matter $(N_2-N_3)$ and $(N_2-N_1)5/2$, the diagonal band shows the
region where $\alphasmz$ is within 2-$\sigma$ of its measured value.
Points along the  vertical lines give the same value of $\mgut$, and
the lines corresponding to $\mgut =10^{15}~\gev$ and $10^{19}~\gev$
are shown.
The black dot indicates the solution
with only fermions in real irreps and the gray dots indicate solutions
when also new scalar particles are added.}
\label{fig1}
\end{figure}

In \fig{fig1} we show the predictions for $\alphasmz$ and
$\mgut$ as functions of $N_2-N_3$ and 
$(N_2-N_1)5/2$, taking $\swsq (\mz )=0.23150\pm 0.00016$
and $\alpha^{-1} (\mz )=128.936\pm 0.0049$~\cite{alt}. 
Agreement with the measured value
$\alphasmz =0.119\pm 0.003$ gives a 
well-defined correlation between $N_2-N_3$ and 
$(N_2-N_1)5/2$, shown by the diagonal band in \fig{fig1}. The requirement 
that unification is achieved 
($\agut >0$) in the perturbative domain ($\agut <1$) imposes the 
constraint
\beq
2N_2-N_1 \lsim 31 -4(N_g-3).
\eeq


A further limitation of the available parameters comes from the request that
$\mgut$ is sufficiently smaller than the Planck mass, in order to trust 
field theory without quantum gravity, and sufficiently large to avoid 
quick proton decay. GUT gauge bosons with mass $\mgut$ mediate the decay
$p\to \pi^0 e^+$ with lifetime
\bea
\tau (p\to \pi^0 e^+)&=&
\frac{4~f_\pi^2~\mgut^4}{\pi m_p\agut^2(1+D+F)^2\alpha_N^2
[A_R^2+(1+|V_{ud}|^2)^2A_L^2]}\\
&=&
\left( \frac{\mgut}{10^{16}~\gev}\right)^4 \left( \frac{1/35}{\agut}
\right)^2\left(\frac{0.015~\gev^3}{\alpha_N}\right)^2
\left(\frac{5}{A_L}\right)^2
~4.4\times 10^{34}~
{\rm yr}.
\eea
Here $f_\pi =131$~MeV, the chiral Lagrangian factor is $(1+D+F)=2.25$, 
and we have taken the operator
renormalization factors $A_L=A_R=5$. For the hadronic
matix element $\alpha_N$, we take the lattice 
result~\cite{lattice} $\alpha_N=0.015~\gev^3$.
From the Super-Kamiokande
limit~\cite{kam} $\tau (p\to \pi^0 e^+)>5.3\times 10^{33}$~yr, we obtain
\beq
\mgut >  \left(\frac{\agut}{1/35}\right)^{1/2}
\left(\frac{\alpha_N}{0.015~\gev^3}\right)^{1/2}
\left(\frac{A_L}{5}\right)^{1/2}
~6\times 10^{15}~\gev .
\label{limitproton}
\eeq

The measured value of $\alphasmz$ and the upper and lower bounds on 
$\mgut$ (given by the vertical lines in \fig{fig1})
select an allowed region in the space of the Dynkin indices of new matter.
From this result,
we can already conclude that one-step unification cannot be achieved by 
adding to the SM only particles carrying no colour charges. Indeed, for
$N_3=0$ we find an upper bound on $\mgut$, which is maximized for 
$N_1=0$, corresponding to $\mgut <2\times 10^{14}~\gev$. This bound becomes
much more stringent if the new weak particles carry hypercharge. At any rate,
the upper bound on $\mgut$ is inconsistent with \eq{limitproton}. Also 
from \fig{fig1} it is apparent that there is no solution when $N_2=0$.
Therefore unification requires a new set of particles carrying both 
colour and weak charges.

Since the variables $N_i$ can only take discrete values, only certain points
in the region of \fig{fig1} are physical. To procede, we elaborate 
on the new particle content.  We start by considering only
fermions in real irreps, as this appears to be the most interesting 
physical case. Fermions
in chiral representations, from incomplete GUT multiplets, generically
introduce anomalies. Moreover, they can only get mass from electroweak
symmetry breaking, and they are typically ruled out either by direct
experimental searches or by their virtual effects on precision measurements.
New light scalars, besides the Higgs, imply further fine-tunings and therefore,
for simplicity, it is preferable to exclude them. At any rate, we will later   
generalize our results to the case of chiral fermions and scalars.

Since eventually we want to unify
the theory, the new particles should belong to some (incomplete) GUT multiplet.
This restricts the allowed quantum numbers. If, for simplicity, we
consider the fermions to be part of 
the $5+\bar 5$, $10+\bar {10}$, $15+\bar {15}$, $24$
irreps of $SU(5)$ and exclude higher dimensional irreps, the possible new
particles are
\bea
Q=(3,2,1/3)+({\bar 3},2,-1/3)&~~~~&U=(3,1,4/3)+({\bar 3},1,-4/3)
\nonumber\\
D=(3,1,-2/3)+({\bar 3},1,2/3)&~~~~&L=(1,2,1)+(1,2,-1)
\nonumber\\
E=(1,1,2)+(1,1,-2)           &~~~~&V=(1,3,0)
\nonumber\\
G=(8,1,0)&~~~~&X=(3,2,-5/3)+({\bar 3},2,5/3)
\nonumber\\
T=(1,3,2)+(1,3,-2)&~~~~&S=(6,1,-4/3)+({\bar 6},1,4/3).
\eea
Introducing a generic number $n_a$ of multiplets belonging to the 
irreps $a$ listed above, we find that their contributions to the 
$\beta$-function are
\bea
N_2-N_3&=&2A \label{n23}\\
\frac{5}{2}(N_2-N_1)&=&2(A+3B)\label{n21}\\
N_2&=&2(3n_Q+n_L+2n_V+3n_X+4n_T)
\eea
\bea
A&\equiv &n_Q-n_U-n_D+n_L+2n_V-3n_G+n_X+4n_T-5n_S
\label{eqa}\\
B&\equiv &2n_Q-n_U-n_E+n_V+n_G-2n_X-n_T-n_S\label{eqb}
\eea
Therefore, the allowed points of the plane in \fig{fig1} are those
obtained from eqs.~(\ref{n23}) and (\ref{n21}) for integers (positive or
negative) values of $A$ and $B$. 
This is actually true independently of the fermion content of the theory,
provided that the fermion representation is real and the new particles
belong to some (incomplete) $SU(5)$ multiplet (see footnote \ref{obs}). 

Negative values of $A$ correspond to solutions where $\mgut >M_{\rm Pl}$,
which we discard. For $A=1$, we find a solution with $B=1$, corresponding
to $\alphasmz =0.120$ and $\mgut =9.8 \times 10^{14}$. The minimal particle
content is given by $(E+Q)$ or $(D+V)$. The case $(D+V)$ is particularly 
interesting, because $V$ contains a dark matter candidate, linking the 
new particles to the TeV scale. However, even though $\alphasmz$ is in
agreement with measurements, $\mgut$ is in conflict with the 
proton-decay bound in \eq{limitproton}. 
We will discard this solution.

We are left only with the case $A=0$. The solution for $B=2$ gives
$\alphasmz=0.102$ and $\mgut =1.6 \times 10^{16}$. The value of $\mgut$
is consistent with \eq{limitproton}, but $\alphasmz$ is rather low and a 
certain amount of threshold effects are beneficial, as we will discuss 
in the next section. 
Possible particle
contents can be obtained by solving eqs.(\ref{eqa}) and (\ref{eqb}). 
The minimal
options
are $(Q+D)$, $(L+V+G)$, $(V+2G+T)$, $(U+2V+G)$, $(2Q+2U)$, or
$(Q+2D+L)$. Those containing $L$ or $V$ have a dark matter candidate.
The case $(Q+D)$ has been proposed in ref.~\cite{wagner} as a cure
to the discrepancy between measurements and SM prediction of the b-quark
forward-backward asymmetry. The particle content of $(L+V+G)$ corresponds
to the supersymmetric SM with a single Higgs doublet: it is thus the
case advocated in ref.~\cite{savas}.
Indeed, we can identify $L$ with the (Dirac)
higgsino components, interpret $V$ and $G$ as the W-ino
and gluino, and add a gauge singlet B-ino (which does not affect the
gauge-coupling evolution at one-loop)
to complete the gaugino spectrum.

The generalization of these results is now straightforward. Chiral fermions
can contribute to $A$ and $B$ with half-integers and complex scalars with
multiples of $1/4$. The solutions with acceptable $\mgut$ and $\alphasmz$
(within 2-$\sigma$) are collected in table~1. 
The explicit 
particle content of the solution listed in table~1
can be obtained from  eqs.~(\ref{eqa}) and (\ref{eqb}).
Notice that
the case $(N_2-N_3)=1/2$ and $(N_2-N_1)5/2=25/2$ corresponds to the usual
supersymmetric extension of the SM with two Higgs doublets.
From the results in table~1 we observe that some of the solutions
have a value of $\mgut$ close to the experimental limit and therefore predict
a measurable proton-decay rate.
Two-loop 
renormalization group and threshold effects can give shifts of $\alphasmz$
even larger than the experimental error. These effects split the degeneracy 
of the various models, since they depend on the matter content through more
parameters than simply $(N_2-N_3)$ and $(N_2-N_1)$. For instance, in the 
minimal supersymmetric model, two-loop running effects amount to a shift
$\Delta  \alphasmz =+0.011$. Low-energy threshold corrections depend on
the detailed mass spectrum. The contributions from individual particles can
be very significant, although in the minimal supersymmetric model a certain
amount of cancellation among the various contributions typically occurs.

\begin{table}
\label{table1}
\centering
\begin{tabular}{|c|c|c|c|c|c|}
\hline\hline
$A$ & $B$ &$N_2-N_3$& $\frac{5}{2}(N_2-N_1)$ & $\alphasmz$ & $\mgut$ [GeV] \\
\hline
$-\frac{1}{2}$ &$\frac{13}{4}$&$-1$ &18.5 &0.121 &$6.2\times 10^{18}$\\
$-\frac{1}{4}$ &$\frac{11}{4}$& $-0.5$ & 16 &0.115 & $5.0\times 10^{17}$\\
 0 & $\frac{5}{2}$&           0 & 15 & 0.120 &$2.0\times 10^{17}$\\
$\frac{1}{4}$& 2&0.5 & 12.5 &0.115 &$2.4\times 10^{16}$\\
$\frac{1}{2}$ &$\frac{7}{4}$&1& 11.5 &0.120 &$1.1\times 10^{16}$\\
$\frac{3}{4}$ &$\frac{5}{4}$&1.5 &9 &0.115 &$1.9\times 10^{15}$\\
\hline\hline
\end{tabular}
\caption{{$N_{1,2,3}$ are the Dynkin indices of the particle content of new 
matter (including scalars) necessary to achieve gauge-coupling unification 
in the SM. The corresponding predictions for $\alphasmz$ and $\mgut$ are
also listed. The values of $A$ and $B$ and eqs.~(\ref{eqa}) and 
(\ref{eqb}) determine the particle multiplets of new matter.
}}
\end{table}

In summary, we have classified the particle content of new matter at the
electroweak scale, necessary to achieve gauge-coupling unification 
in the SM (see also ref.~\cite{altri}). 
We have found several class of solutions. However, if we
require the presence of only fermions in real irreps, there is only one
class of solutions. Although there is a large degeneracy of models 
with the same values of $(N_2-N_3)$ and $(N_2-N_1)$, therefore belonging
to the same class, the case of Split Supersymmetry proposed in 
ref.~\cite{savas} is the case of minimal field content with a dark matter
candidate.

\subsection{Split Supersymmetry versus Non-Supersymmetric Solutions}
\label{versus}

It is likely that supersymmetry plays a r\^ole in a consistent theory of
gravity, like superstrings. But, if we abandon the naturalness
problem, the rationale for low-energy supersymmetry is lost, since the
preferred superstring vacuum may well completely break supersymmetry.
Nevertheless, we want to argue that the case of Split Supersymmetry 
looks particularly intersting, when compared with
the other non-supersymmetric solutions to gauge unification
that we have found in the previous section.

In general, once we introduce new particles at the weak scale to obtain
gauge-coupling unification, we encounter several problems.
Split Supersymmetry seems to have partial answers to several of them.
Let us briefly analyse them.
 

{\it (i) Splitting of GUT irreps.} Gauge-coupling unification requires
the addition of new incomplete GUT multiplets. It is not always natural
in a GUT to obtain the appropriate content of light particles. 
This is nothing else
than the old doublet-triplet splitting problem
for the Higgs multiplet. 
In Split Supersymmetry, once the necessary Higgs doublet-triplet
splitting is achieved, no other incomplete matter multiplets 
are necessary. Supersymmetry
plays a crucial r\^ole in relating the content of
gauginos to the incomplete irrep of gauge bosons, and higgsinos to Higgs. 

{\it (ii) Light fermions.} To maintain vector-like fermions light, the theory
should posess some
approximate global symmetries. 
Supersymmetry has the advantage of having some of this 
symmetries already built in. Gaugino and higgsino masses can be protected,
even after soft supersymmetry breaking, by an $R$-symmetry and a PQ symmetry,
related to the origin of the $\mu$ term~\cite{giu}. 
We will show that actually, in
the effective theory, only a linear combination of the two symmetries survives,
but this is sufficient to protect gaugino and higgsino masses.

{\it (iii) Existence and stability of dark matter.} In this framework, dark
matter provides the link with the electroweak scale. Therefore the existence
and stability of a weakly-interacting neutral particle is a crucial ingredient.
Supersymmetry
has these features built in, because of the $R$-parity, which can be a 
low-energy consequence of a GUT gauge symmetry.

{\it (iv) Instability of coloured particles.} We have seen that the
condition for one-step unification and proton stability requires the existence
of new coloured particles at the electroweak scale. If these particles 
are stable,
they could be present today, bound in nuclei, and they 
would appear as anomalously heavy isotopes. Their relic
abundance is quite uncertain, because it depends on the mechanism of
hadronization and of nuclear binding~\cite{baer}. However, even allowing for
the most conservative estimates, stable coloured particles are ruled out
by searches for heavy hydrogen isotopes, which excludes that their
number per nucleon is larger than $10^{-28}$ for a mass up to
1~TeV~\cite{lim1} and  $10^{-20}$  up to 10~TeV~\cite{lim2}. The decay of the
new coloured particles could arise from 
mixing with ordinary quarks through renormalizable interactions. However
this mixing typically
introduces unwanted flavour violations, and it is tightly constrained.
If the decay occurs through a non-renormalizable 4-fermion interaction
with scale $\Lambda$,
the lifetime of the coloured particle with mass $M$ is
\beq
\tau \simeq (4\pi)^3\frac{\Lambda^4}{M^5}\simeq 
\left( \frac{\tev}{M}\right)^5 
 \left( \frac{\Lambda}{10^{13}~\gev}\right)^4 ~0.4~{\rm Gyr}.
\eeq
Thus, the decay of the coloured particle requires either a small parameter
or a new threshold at an intermediate scale smaller than about $10^{13}~\gev$,
which can modify the gauge-coupling
evolution. In Split Supersymmetry, the intermediate threshold is provided
by the squark and slepton masses, which mediate gluino decay. 
This threshold consists of a complete GUT irrep (apart from the heavy
Higgs doublet) and therefore does not (much) affect the unification 
condition.
As long as the mechanism of supersymmetry breaking explains the existence
of two widely separated scales, the gluino stability does not pose a 
problem.

Because of the interesting features of Split Supersymmetry,
we proceed to investigate it in more detail.

\section{The Structure of Split Supersymmetry}
\label{sec3}

The spectrum of Split Supersymmetry contains the
higgsino components ${\tilde H}_{u,d}$,   
the gluino ($\tilde g$), the W-ino ($\tilde W$) 
and B-ino
($\tilde B$), and the SM particles with one Higgs doublet.
The most general renormalizable
Lagrangian with a matter parity, besides
gauge-invariant kinetic terms, is given by
\bea
{\cal L}&=&m^2 H^\dagger H-\frac{\lambda}{2}\left( H^\dagger H\right)^2
-\left[ h^u_{ij} {\bar q}_j u_i\epsilon H^* 
+h^d_{ij} {\bar q}_j d_iH
+h^e_{ij} {\bar \ell}_j e_iH \right. \nonumber \\
&&+\frac{M_3}{2} {\tilde g}^A {\tilde g}^A
+\frac{M_2}{2} {\tilde W}^a {\tilde W}^a
+\frac{M_1}{2} {\tilde B} {\tilde B}
+\mu {\tilde H}_u^T\epsilon {\tilde H}_d \nonumber \\
&&\left. +\frac{H^\dagger}{\sqrt{2}}\left( \gtilu \sigma^a {\tilde W}^a
+\gtilup {\tilde B} \right) {\tilde H}_u
+\frac{H^T\epsilon}{\sqrt{2}}\left(
-\gtild \sigma^a {\tilde W}^a
+\gtildp {\tilde B} \right) {\tilde H}_d +{\rm h.c.}\right] ,
\label{lagr}
\eea
where $\epsilon =i\sigma_2$.

The Lagrangian in \eq{lagr} describes the effective theory obtained by
removing squarks, sleptons, charged
and pseudoscalar Higgs from the supersymmetric SM. These states are
assumed to be heavy and, for simplicity, we will assume them to be
all degenerate with mass $\tilde{m}$. The coupling constants of the
effective theory
at the scale $\tilde{m}$ are obtained by matching the
Lagrangian in \eq{lagr}
with the interaction terms of the supersymmetric Higgs doublets
$H_u$ and $H_d$,
\bea
{\cal L}_{\rm susy}&=&
-\frac{g^2}{8}\left( H_u^\dagger \sigma^a H_u + H_d^\dagger \sigma^a H_d
\right)^2
-\frac{g^{\prime 2}}{8}\left( H_u^\dagger H_u - H_d^\dagger  H_d
\right)^2 \nonumber \\
&&+\lambda^u_{ij}H_u^T\epsilon {\bar u}_i q_j
-\lambda^d_{ij}H_d^T\epsilon {\bar d}_i q_j
-\lambda^e_{ij}H_e^T\epsilon {\bar e}_i \ell_j
\nonumber \\
&&-\frac{H_u^\dagger}{\sqrt{2}}\left( g \sigma^a {\tilde W}^a
+g^\prime {\tilde B} \right) {\tilde H}_u
-\frac{H_d^\dagger}{\sqrt{2}}\left(
g \sigma^a {\tilde W}^a
-g^\prime {\tilde B} \right) {\tilde H}_d +{\rm h.c.}
\label{lagrs}
\eea
Once the Higgs doublet $H=-\cos\beta \epsilon H_d^*+\sin\beta H_u$
is fine-tuned to have small mass term, the matching conditions 
of the coupling constants in \eq{lagr} at the scale $\tilde{m}$ are
obtained by replacing $H_u\to \sin\beta H$, $H_d\to \cos\beta \epsilon
H^*$ in \eq{lagrs}:
\bea
\lambda(\tilde m )& =& \frac{\left[ g^2(\tilde m )+g^{\prime 2}(\tilde m )
\right]}{4} \cos^22\beta ,
\label{condh}\\
h^u_{ij}(\tilde m )=\lambda^{u*}_{ij}(\tilde m )\sin\beta , &&
h^{d,e}_{ij}(\tilde m )=\lambda^{d,e*}_{ij}(\tilde m )\cos\beta ,\\
\gtilu (\tilde m )= g (\tilde m )\sin\beta ,&&
\gtild (\tilde m )= g (\tilde m )\cos\beta ,\\
\gtilup (\tilde m )= g^\prime (\tilde m ) \sin\beta ,&&
\gtildp (\tilde m )= g^\prime (\tilde m )\cos\beta . 
\label{condg}
\eea

In the context of a supersymmetric theory it is possible to argue
that the gaugino masses $M_i$ or the higgsino mass $\mu$ are much
smaller than the typical scale
because they are protected by an $R$-symmetry and a PQ symmetry,
respectively. However, the symmetries of the effective Lagrangian
in \eq{lagr} are not enhanced if we set $M_i=0$ or $\mu =0$, separately.
This is because supersymmetry has been explicitly broken by eliminating
the scalar fields. Nevertheless, if we simultaneously set
$M_i=\mu =0$, the effective theory remains invariant under the product of 
an $R$-symmetry (with $R$-charges $R[H_u]=2$, $R[H_d]=0$) and
a supersymmetric PQ symmetry (with charges $PQ[H_u]=PQ[H_d]=-1$). 
In other words,
\eq{lagr} is invariant under a global $U(1)$ symmetry with charges
$S[{\tilde B}]=S[{\tilde W}]=S[H]=-S[{\tilde H}_d]/2$, 
$S[{\tilde H}_u]=0$, and with quarks and leptons with the appropriate
charges. This symmetry is spontaneously broken by the Higgs vev and explicitly
broken by $M_i$ or $\mu$. 

This result shows that the choice of keeping only light gauginos or 
light higgsinos, but not both of them simultaneously, is not radiatively 
stable. It is interesting that the only consistent choice (forgetting
the tuning of the Higgs mass) of splitting the supersymmetric spectrum
is the one that successfully reproduces gauge-coupling unification.
Another consequence of this result is that the $\mu$ parameter mixes
with $M_{1,2}$ under renormalization effects,
as discussed in sect.~\ref{seccou}.

A particular situation occurs when $\tan\beta \to \infty$, since
$h^d,~h^e,~\gtild~,\gtildp$ vanish in this limit. Then
\eq{lagr}, for $\mu =0$, has an additional global symmetry (with
charges ($[{\tilde B}]=[{\tilde W}]=0$, $[H]=[{\tilde H}_u]= [{\tilde H}_d]$).
When we set $M_i=0$ and keep $\mu$ non-vanishing, we find a different
global symmetry (with
charges ($[{\tilde B}]=[{\tilde W}]=[H]/2=[{\tilde H}_u]= -[{\tilde H}_d]$).  
Therefore, we expect no renormalization mixing between $M_{1,2}$ and
$\mu$ in the limit of large $\tan\beta$.

\subsection{Unification}

To make a precise assessment on gauge-coupling constant unification in 
Split Supersymmetry, it is necessary to study the 2-loop renormalization group
evolution, including one-loop threshold effects. Between the unified scale
$\mgut$ and the scale of heavy scalars $\tilde{m}$, the theory is exactly
supersymmetric. Below the scale $\tilde{m}$, we use the spectrum of
Split Supersymmetry, with gauginos and higgsinos included in the 2-loop
evolution. Then we include separate threholds for the gluino $M_3$ and
for weak gaugino and higgsino masses, and properly evolve between the 
two scales. 
Thresholds of the top quark and SM gauge bosons are taken into account in the
usual way~\cite{sogl}. The relevant renormalization-group equation are given
in the appendix. We use the same SM input values used in sect.~\ref{sec2}.

\begin{figure}
\begin{center}
\epsfig{file=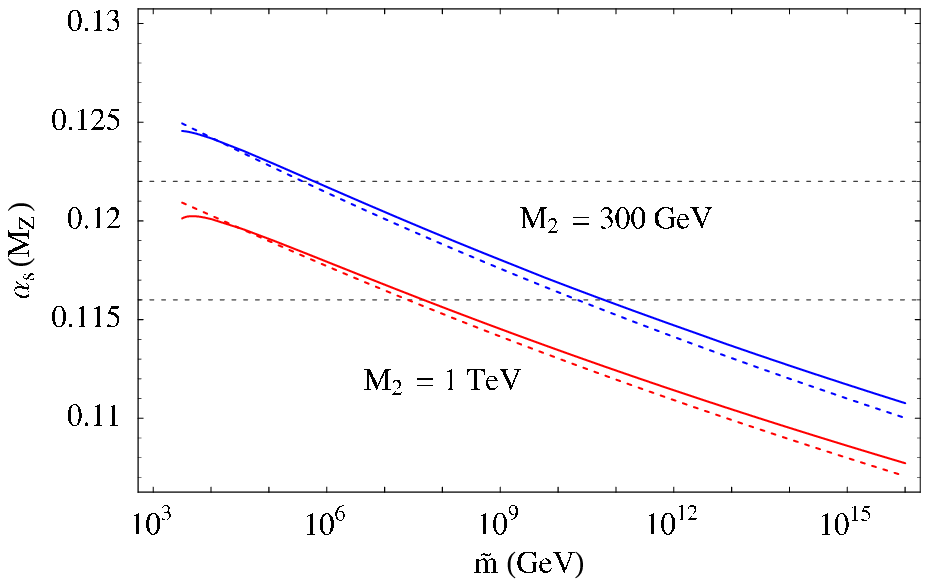,width=0.6\textwidth}
\end{center}
\caption{The unification 
prediction for $\alphasmz$ as a function of $\tilde m$. The 
solid line corresponds to $\tan\beta =50$ and the dashed line to
$\tan\beta =1.5$. The gaugino and higgsino thresholds are taken by assuming
gaugino mass unification, and $\mu =M_2$. The results for $M_2 =300~\gev$ and
1~TeV are shown.}
\label{gauge}
\end{figure}

The prediction of $\alphasmz$, as a function of
the intermediate scale $\tilde{m}$,  
is shown in \fig{gauge}. We have shown 
our results for values of $\tilde{m}$ as large as $\mgut$, although the
condition that the gluino lifetime is shorter than the age of the universe
requires $\tilde{m}\lsim 10^{13}$~GeV. However, we recall that the limit
from the search for heavy isotopes is valid only up to a gluino mass of
10~TeV.
 
A value of $\tilde{m}$ larger than in conventional supersymmetric
models improves the agreement between the theoretical prediction and
experimental data. However, because of the theoretical uncertainty due
to unknown GUT thresholds and supersymmetric thresholds at the scale
$\tilde{m}$, it is not possible to extract firm bounds on the
parameters. A dependence of $\alphasmz$ on $\tan\beta$ arises from
two-loop effects proportional to the Yukawa couplings and the gaugino
couplings $\tilde{g}$, but the numerical contribution is marginal. On
the other hand, the effect of the gaugino and higgsino threshold is
important.

\begin{figure}
\begin{center}
\epsfig{file=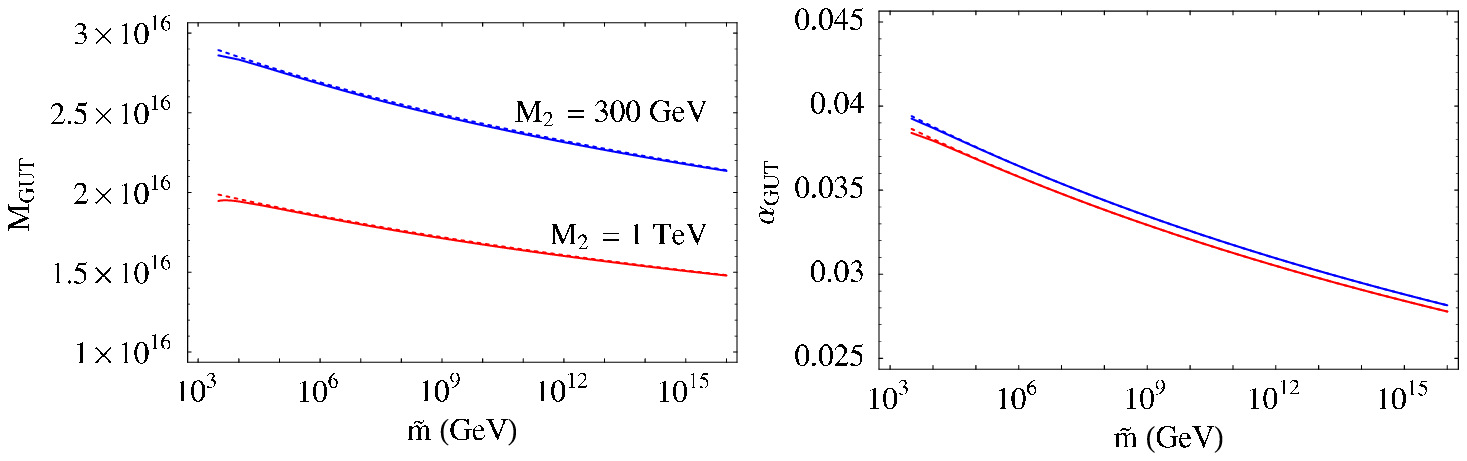,width=\textwidth}
\end{center}
\caption{The unification prediction for $\mgut$ and $\agut$ as functions 
of $\tilde m$. The 
solid line corresponds to $\tan\beta =50$ and the dashed line to
$\tan\beta =1.5$. The gaugino and higgsino thresholds are taken by assuming
gaugino mass unification, and $\mu =M_2$. The results for $M_2 =300~\gev$ and
1~TeV are shown.}
\label{masgut}
\end{figure}

The prediction for $\mgut$ is shown in \fig{masgut}. The unified mass
decreases as $\tilde{m}$ grows, but the proton-decay rate from dimension-6
operators remains unobservably small, at present. We recall that proton
decay through dimension-5 operators is suppressed by the large squark mass.
The value of $\agut$ decreases with $\tilde{m}$, because of the smaller
particle content of Split Supersymmetry with respect to the ordinary case,
as shown in \fig{masgut}.

Before discussing the unification of bottom and $\tau$ masses, let us 
discuss the running of the top Yukawa coupling. It is well known that in 
the MSSM the Landau pole sets an upper limit on the top coupling, which 
translates in a lower limit on $\tan\beta$. In split supersymmetry, the 
Landau pole is met before the unification scale for lower values of 
$\tan\beta$. This happens for two reasons. First, below $\tilde m$,
the Yukawa coupling belongs to a theory with a single Higgs doublet
and it is 
smaller by a factor $\sin\beta$ than in the two-Higgs theory. Second, 
the contribution of the top coupling to its evolution is smaller (see
equations in the 
appendix). As a consequence, the value of the coupling at the matching 
with the supersymmetric theory is smaller, and this allows a smaller value 
of $\tan\beta$. Such low values of $\tan\beta$ are not necessarily 
ruled out in this scenario, as the Higgs mass can be made sufficiently 
large even for low values of $\tan\beta$ (see sect.~\ref{sechig}).

\begin{figure}
\begin{center}
\epsfig{file=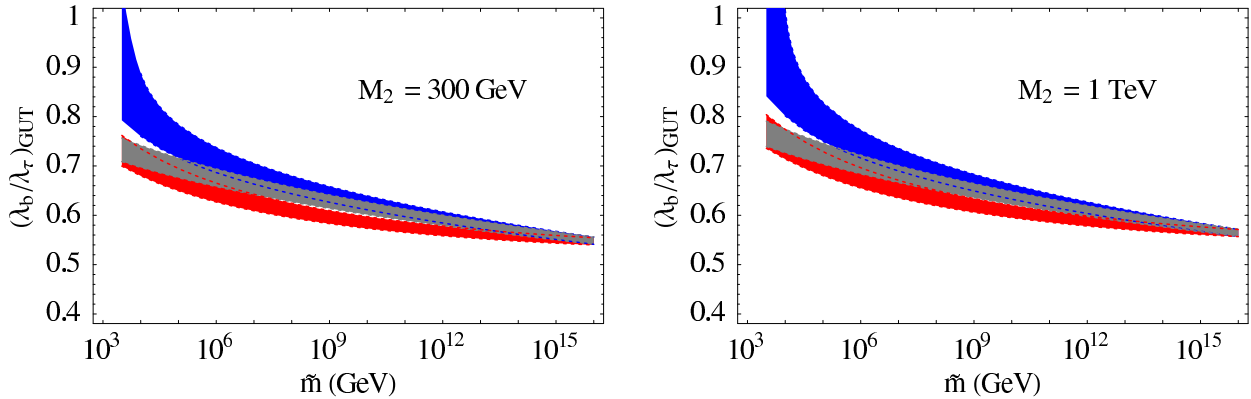,width=\textwidth}
\end{center}
\caption{The Yukawa-coupling ratio $\lambda_b/\lambda_\tau$ evaluated
at $\mgut$. The gaugino and higgsino thresholds are taken by assuming
gaugino mass unification, and $\mu =M_2$. The results for $M_2 =300~\gev$ and
1~TeV are shown. The bands correspond to 1-$\sigma$ uncertainties
in $m_b$ and $m_t$, and $\tan\beta =1.5,10,50$, from top to bottom.
}
\label{btau}
\end{figure}

Since the model has a successful gauge-coupling unification, it is 
natural  to study if also the bottom and $\tau$ masses can unify at the 
same scale. We take the bottom-quark $\overline{\rm MS}$ running mass 
$m_b(m_b)=4.2\pm 0.1$~GeV. Although the effective theory has only one 
Higgs doublet, the ratio $m_b/m_\tau$ depends on $\tan\beta$ because of 
the running above $\tilde{m}$. In \fig{btau} we show the Yukawa-coupling 
ratio $\lambda_b/\lambda_\tau$ at the unification scale, as a function 
of $\tilde{m}$, for different values of $\tan\beta$. Here the trend with 
$\tilde{m}$ is the opposite than for the prediction of $\alphasmz$, and 
heavy scalars make $b$--$\tau$ unification more difficult for a given 
$\tan\beta$. For large $\tan\beta$ this is also because in Split 
Supersymmetry one can not rely on sizable finite corrections at large 
$\tan\beta$~\cite{ratt}. Since the coefficient $A$ of the trilinear soft 
terms is forbidden by the same $R$-symmetry that protects gaugino 
masses, the large-$\tan\beta$ corrections are suppressed by powers of 
the heavy-scalar masses. 
For small $\tan\beta$, one obtains lower values of 
$b$--$\tau$, as a consequence of the slower running of the 
top Yukawa coupling discussed above (the top Yukawa increases the
value of the bottom mass and does not affect the tau mass). 

Despite heavy scalars make $b$--$\tau$ unification more difficult
\emph{at a given $\tan\beta$}, the prospects for $b-\tau$ unification
at low $\tan\beta$ in Split Supersymmetry are actually better than in
the MSSM. In fact, in both cases the top Yukawa enhancement of $m_b$
can be made large enough by using values of $\tan\beta$ close enough
to the Landau pole lower limit (as in the MSSM, this involves a degree
of fine-tuning). In the MSSM, such values of $\tan\beta$ are excluded
by the Higgs mass limit. Because of the slower top running, in Split
Supersymmetry the values of $\tan\beta$ required to enhance $m_b$ are
even lower. However, they are not excluded by the Higgs mass limits,
as we will see later on. Therefore, $b$--$\tau$ unification at low
$\tan\beta$ is not ruled out for heavy scalars, differently than in
the MSSM. Note that we have neglected in our discussion the
possibility of a contribution from neutrino Yukawa
couplings~\cite{neutr} (which enhances the tau mass and therefore goes
in the wrong direction).  Also, the $m_b/m_\tau$ ratio at $\mgut$ can
be enhanced because of contribution from higher Higgs representations
or because of the flavour structure of the Yukawa mass matrices.

\subsection{Gaugino Couplings}
\label{seccou}

A testable prediction of Split Supersymmetry is the deviation of the
equality between gauge and gaugino couplings. This could be detected
by precise measurements of the gaugino production cross section. Such
tests have already been proposed in the case of low-energy 
supersymmetry~\cite{obl}. The effect in Split Supersymmetry is much enhanced
by the heaviness of squarks and slepton. A log resummation is required
to compute the effect and the
relevant renormalization-group equations are given in the appendix.

\begin{figure}
\begin{center}
\epsfig{file=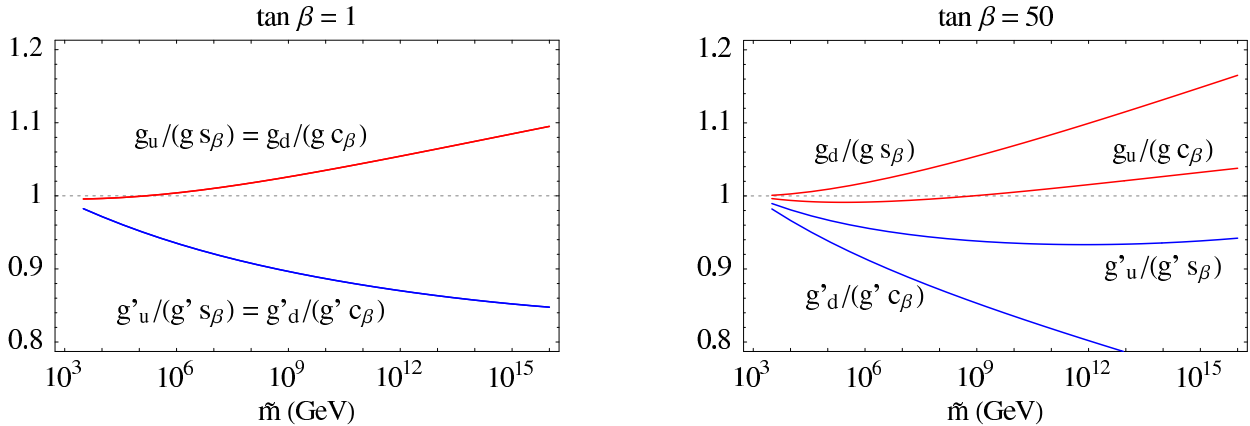,width=\textwidth}
\end{center}
\caption{
  The gaugino couplings in units of the gauge couplings $\gtilu
  /(g\sin\beta)$, $\gtild /(g\cos\beta) $, $\gtilup /(g^\prime\sin\beta)$, 
  $\gtildp /(g^\prime \cos\beta) $ as functions of $\tilde m$,
  calculated at the weak scale. The left frame corresponds to
  $\tan\beta =1$ and the right one to $\tan\beta =50$. }
\label{ggau}
\end{figure}

The gaugino couplings $\tilde{g}$, which satisfy the boundary
conditions in \eq{condg} at the scale $\tilde{m}$, are evolved to the
weak scale.  Numerical results are shown in \fig{ggau}.  To understand
the behaviour, it is useful to take the analytic solution of the
renormalization-group equations, in the limit of small
$\ln(\tilde{m}/\bar \mu )$, where $\bar \mu$ is the renormalization
scale: 
\bea 
\left. \frac{\gtilu }{g\sin\beta}\right|_{\bar \mu}
&\simeq &1+\frac{\ln(\tilde{m}/\bar \mu )} {(4\pi)^2}\left[
\left( \frac{13}{3} +\frac{7}{4}\cos^2\beta \right) g^2
-\frac{3}{4}\cos^2\beta g^{\prime 2}
-3h_t^2\right]_{
  \tilde{m}} \label{gup}\\
\left. \frac{\gtilup }{g^\prime \sin\beta}\right|_{\bar \mu} &\simeq
&1+\frac{\ln(\tilde{m}/\bar \mu )} {(4\pi)^2}\left[ 
-\frac{9}{4}\cos^2\beta g^2
+\left( 7-\frac{3}{4}\cos^2\beta \right)g^{\prime 2}
-3h_t^2\right]_{ \tilde{m}} .\label{gupp} 
\eea
The equations for $\gtild /(g\cos\beta) $ and $\gtildp
/(g^\prime\cos\beta) $ are obtained from \eq{gup} and \eq{gupp},
respectively, by replacing $\cos 2\beta \to -\cos 2\beta$ in the
right-hand side.  Gauge and top-quark Yukawa contribute to the running
of the gaugino couplings with opposite signs. In $\gtilu /(g
\sin\beta)$ and $\gtild /(g\cos\beta) $, the $SU(2)$ contribution has
a very large coefficient which overtakes the top-Yukawa effect. In
$\gtilup /(g^\prime\sin\beta) $ and $\gtildp /(g^\prime\cos\beta) $,
the most important gauge effects come from hypercharge, and top-quark
effects tend to dominate. A significant cancellation of the two
competing effects can occur in $\gtilup /(g^\prime \sin\beta) $ or
$\gtilu /(g \sin\beta) $.

\subsection{Gaugino Masses}
\label{secgmas}

\begin{figure}
\begin{center}
\epsfig{file=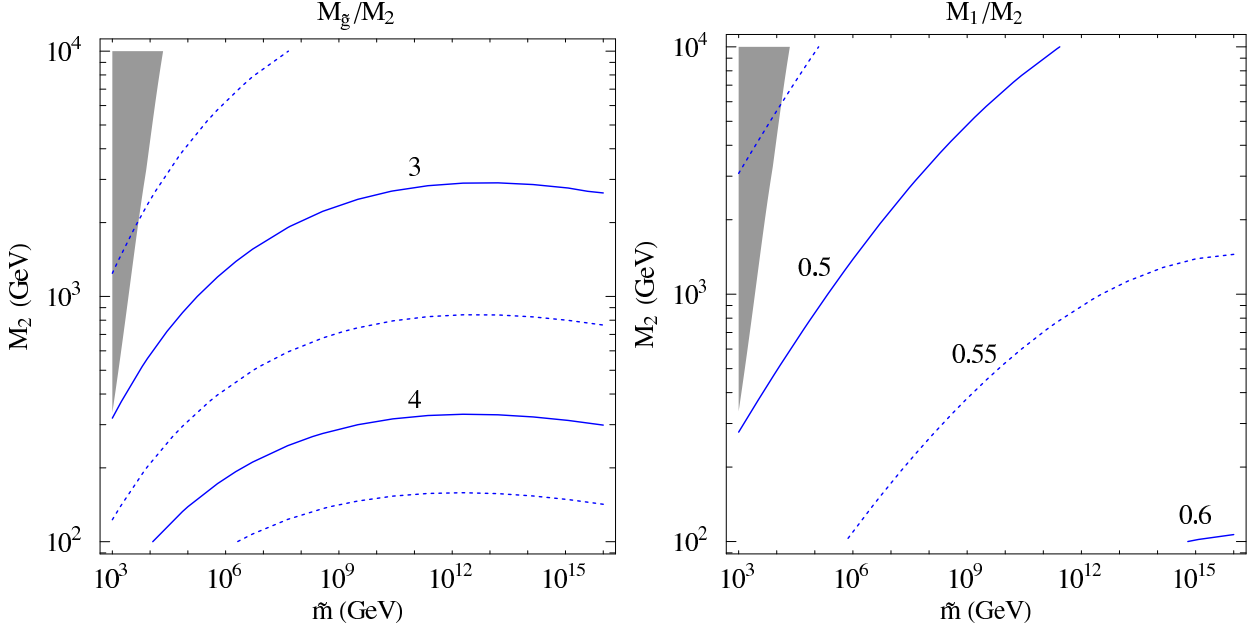,width=\textwidth}
\end{center}
\caption{
  The ratios of the pole gluino mass $M_{\tilde g}$ to $M_2$ (left
  frame) and $M_1/M_2$ (right frame) as functions of $\tilde m$ and
  $M_2$, assuming gaugino mass unification. We have taken a large
  value of $\tan\beta$, so that the mixing between $M_{1,2}$ and $\mu$
  is irrelevant. The shaded region corresponds to $\tilde m <
  M_{\tilde g}$.}
\label{m3m2}
\end{figure}

We have computed the gluino pole mass with two-loop
renormalization-group evolution and threshold effects, with the
equations given in the appendix. To help the discussion, consider the
one-loop expressions of the gaugino mass at the scale $\bar \mu$ (in
the limit of large $\tan\beta$) \bea M_3(\bar \mu )&=&M_3(\mgut )\frac
{\alpha_s (\tilde{m})}{\agut}
\left[ \frac {\alpha_s (\bar \mu )}{\alpha_s (\tilde{m})}\right]^{9/5},\\
M_2(\bar \mu )&=&M_2(\mgut )\frac {g^2 (\tilde{m})}{4\pi\agut}
\left[ \frac {g^2 (\bar \mu )}{g^2 (\tilde{m})}\right]^{33/7},\\
M_1(\bar \mu )&=&M_1(\mgut )\frac {5g^{\prime 2}
  (\tilde{m})}{12\pi\agut} \left[ \frac {g^{\prime 2} (\bar \mu
    )}{g^{\prime 2} (\tilde{m})}\right]^{1/15}.  \eea The
renormalization of the gaugino masses below $\tilde{m}$ is
substantial. However, the ratio between gluino and weak-gaugino masses
$M_{\tilde g}/M_2$ and the ratio $M_2/M_1$ do not significantly depend
on $\tilde{m}$, as shown in \fig{m3m2}.  This can be explained because
the leading-logarithm evolution of the ratio $M_i/M_j$ is proportional
to the difference between the numerical coefficients of the $M_i$ and
$M_j$ anomalous dimensions. Therefore, intermediate thresholds of
complete GUT multiplets (like squarks and sleptons at the scale
$\tilde{m}$) have limited effect on the renormalization of the ratio
$M_i/M_j$.

\begin{figure}
\begin{center}
\epsfig{file=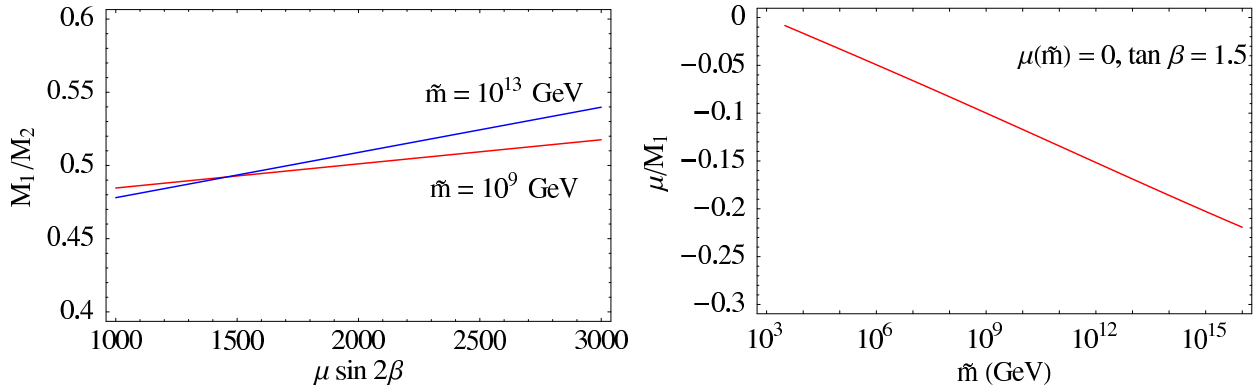,width=\textwidth}
\end{center}
\caption{Left frame: the variation of the $M_1/M_2$ ratio with $\mu\sin 
2\beta$. Right frame:
The value of the $\mu$ parameter at the weak scale in units of
the smallest gaugino mass parameter $M_1$, assuming $\mu =0$ at the scale
$\tilde m$, and gaugino mass unification. We have taken $\tan\beta
=1.5$, since low values of $\tan\beta$ enhance the effect of radiative corrections.}
\label{m1m2mu}
\end{figure}

As explained above in terms of symmetries, we expect the unusual
feature of mixing between $\mu$ and $M_{1,2}$ under renormalization effects.
This is confirmed in the equations given in the appendix. For illustrative
purposes, consider the approximate solutions for a renormalization scale 
$\mu$ not much smaller than $\tilde{m}$
\bea
M_2(\bar \mu ) &\simeq& \frac{\alpha M_2(\mgut )}{\swsq\agut}\left[ 
1+\frac{\ln(\tilde{m}/\bar \mu )}
{4\pi}\left( \frac{11 \alpha}{\swsq}- \frac{2\agut\mu\sin 2\beta}{M_2(\mgut )}
\right) \right]_{\tilde{m}} \\
 M_1(\bar \mu ) &\simeq& \frac{\alpha M_1(\mgut )}{\cwsq\agut}\left[ 
\frac{5}{3}-\frac{\ln(\tilde{m}/\bar \mu )}
{4\pi}\left( \frac{5 \alpha}{3\cwsq}+\frac{2\agut\mu\sin 2\beta}
{M_1(\mgut )}
\right) \right]_{\tilde{m}} ,
\eea
where couplings on the right-hand side are evaluated at the scale $\tilde{m}$.
Notice that indeed, as explained above, the mixing disappears for large
$\tan\beta$. Even for $\tan\beta \simeq 1$, the mixing effect is 
suppressed with respect to the renormalization of $M_2$. The numerical
evaluation of the ratio $M_1/M_2$, using the equations given in the appendix,
is shown in \fig{m3m2}. The effect of the mixing with $\mu$ is very limited,
as
exhibited in \fig{m1m2mu}.

Because of the mixing between $M_{1,2}$ and $\mu$, we could imagine to 
entirely generate $\mu$ by radiative effects. Therefore we consider the
possibility that $\mu =0$ at the matching scale $\tilde{m}$, and we
compute the low-energy value, with the results shown in \fig{m1m2mu}.
We find that $\mu$ turns out to be 
significantly smaller than the gaugino masses,
and therefore the lightest neutralino is a higgsino. In the next section,
we will show that a dark-matter higgsino requires $\mu \simeq$~1.0--1.2~TeV.
Therefore, in this case, the whole spectrum is quite heavy.

\begin{figure}
\begin{center}
\epsfig{file=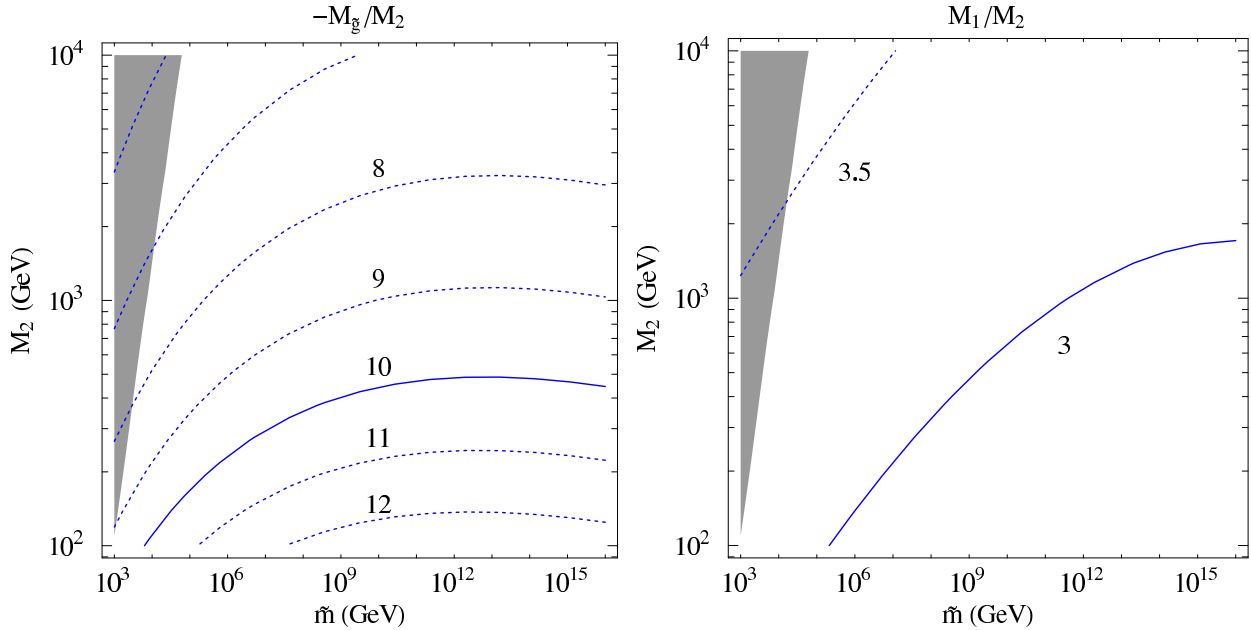,width=\textwidth}
\end{center}
\caption{The ratios of the pole gluino mass $M_{\tilde g}$
to $M_2$ (left frame)
and $M_1/M_2$ (right frame) as functions 
of $\tilde m$, assuming  gaugino mass conditions
from anomaly-mediation. 
We have
taken a large value of $\tan\beta$, so that the mixing between
$M_{1,2}$ and $\mu$ is irrelevant. The shaded region corresponds to
$\tilde m < M_{\tilde g}$.}
\label{m3m2anom}
\end{figure}

So far we have considered a unification condition for gaugino masses. This
appears quite justified, since unification is a crucial ingredient of
this analysis. However, we can consider another interesting possibility.
If supersymmetry breaking occurs in a sector without gauge singlets (as
in models with dynamical symmetry breaking), than the leading contribution
to gaugino masses comes from anomaly mediation~\cite{anom}. The value of the
gaugino masses traces the $\beta$-functions all the way down to the scale
$\tilde{m}$, below which the spectrum is no longer supersymmetric. Therefore
we consider the following
gaugino-mass boundary conditions at the scale $\tilde m$
\beq
M_i = \frac{\beta_{g_i}}{g_i}  m_{3/2} .
\label{anom}
\eeq
Here $\beta_{g_i} $ are the beta-functions of the gauge coupling
$g_i$ (see appendix)
and $m_{3/2}$ is the vev of the auxiliary component of the supergravity
compensator field. The anomaly-mediation mass relations are
modified by the non-supersymmetric running below the scale $\tilde m$.
The ratio of the gaugino masses
with anomaly-mediation boundary conditions are given in \fig{m3m2anom}. 
The W-ino always remains the lightest gaugino.

\begin{figure}
\begin{center}
\epsfig{file=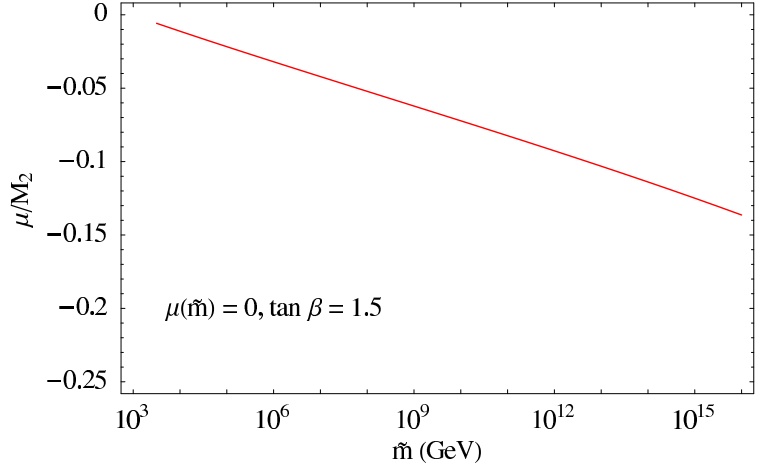,width=0.6\textwidth}
\end{center}
\caption{The value of the $\mu$ parameter at the weak scale in units of
the smallest gaugino mass parameter $M_2$, assuming $\mu =0$ at the scale
$\tilde m$, and anomaly-mediated relations for the  gaugino masses.
We have taken $\tan\beta =1.5$,
which enhances the effect of radiative corrections.}
\label{muanom}
\end{figure}

We have also considered the case of a radiatively-generated $\mu$ parameter
in Split Supersymmetry with anomaly mediation and the result in shown
in \fig{muanom}. As with gaugino unification, also in this case the
higgsino turns out to be the lightest neutralino, but the hierarchy 
between $\mu$ and the gluino mass is amplified by the conditions of
anomaly mediation.

\subsection{Higgs Mass}
\label{sechig}

\begin{figure}
\begin{center}
\epsfig{file=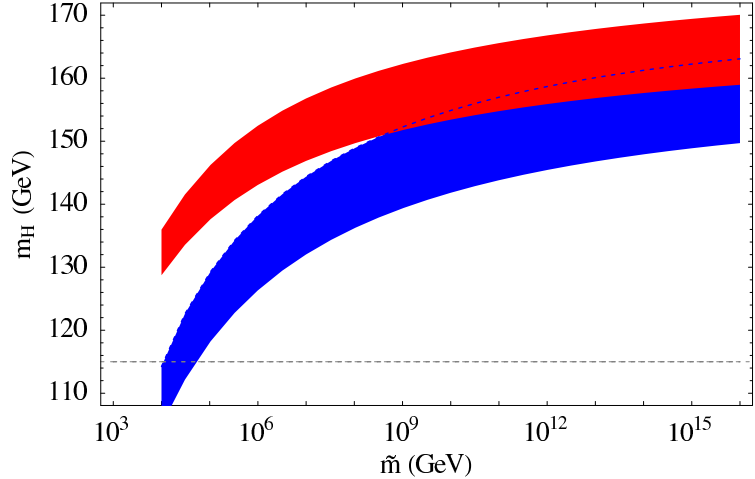,width=0.6\textwidth}
\end{center}
\caption{The value of the Higgs mass
as a function 
of $\tilde m$. The bands include 1-$\sigma$ errors on $m_t$ and
$\alphasmz$. The upper band corresponds to $\tan\beta =50$ and the
lower one to $\tan\beta =1.5$.}
\label{higgs}
\end{figure}

The present lower bound on the SM Higgs mass, $m_H>114.4~\gev$ at 
95\% CL~\cite{hig},  
provides a strong constraint on the parameters of low-energy supersymmetry.
This constraint is relaxed in Split Supersymmetry, because the Higgs
boson mass receives large radiative correction in the evolution from
$\tilde{m}$ to the weak scale. We have solved the relevant equations, contained
in the appendix, and obtained the value of the Higgs mass shown in \fig{higgs}.
The band shows the uncertainty of the predictions due to the 
1-$\sigma$ error on
the top-quark pole mass, $m_t=178.0\pm 4.3~\gev$. The Higgs-mass dependence
on $\tan\beta$ comes primarily from the boundary condition in \eq{condh},
and to a lesser extent from the renormalization-group evolution. 
Because of the large renormalization
effect, values of $\tan\beta$ equal to 1 are allowed, but they require
large values of $\tilde{m}$ and low values of gaugino and higgsino masses.

\subsection{Dark Matter}

As discussed in the introduction, in the absence of the naturalness criterion,
dark matter can provide the link between new physics and the electroweak
scale. It is therefore crucial to study what are the implications of the
request that the lightest neutralino is the dark matter particle.
Differently than in ordinary low-energy supersymmetry, the parameter $\mu$
is not determined by electroweak symmetry breaking, but uniquely by
the relic abundance calculation. In this section, we study this relation.

Let us first consider the case in which the lightest neutralino is 
mostly B-ino. Since squarks and sleptons are decoupled and the B-ino
is a gauge singlet, its only interaction is through its coupling
${\tilde g}^\prime_{u,d}$
with Higgs and higgsinos, given in \eq{lagr}. Therefore, if $\mu \gg M_1$, 
the B-ino is nearly decoupled and it annihilates too weakly in the 
early universe. This means that we need to consider values of
$\mu$ comparable with $M_1$, and thus the lightest neutralino $\chi$ is
always a mixture of gaugino and higgsino and never a pure state. Through
the mixing, this state annihilates efficiently into Higgs and gauge bosons.
The dominant contribution, when $\mu$ is not much larger than $M_1$ comes
from $p$-wave annihilation into longitudinal gauge bosons which,
for $M_1 \gg M_Z$, gives a $\chi$ relic abundance
\beq
\Omega_{\chi}h^2 \simeq 0.1~\frac{\mu^2(M_1^2+\mu^2)^2}{m_\chi^4\tev^2}.
\label{bino}
\eeq

Now we turn to the case in which the lightest neutralino is mainly a higgsino.
The higgsino has gauge interactions which survive in the limit
$M_{1,2}\gg \mu$, and therefore it can be the dark matter, even in a pure
state. Actually the coupling of the lightest higgsino with the $Z$
vanishes when $\mu \gg M_Z$. However, in this limit the other neutral
and charged higgsinos are nearly degenerate in mass (and the lightest state
is neutral) and off-diagonal
couplings of the gauge bosons to higgsinos are allowed. When computing
the relic abundance, it is therefore
important to include the coannihilation of the various channels~\cite{coan}.
When this is done, we find that the relic abundance of a heavy higgsino
in Split Supersymmetry 
is\footnote{The numerical 
calculations of the dark-matter
relic abundances and detection rates presented in this paper
have been performed using the fortran package DarkSUSY~\cite{ds} adapted
to the case of Split Supersymmetry.}
\beq
\Omega_{\tilde H}h^2=0.09~\left( \frac{\mu}{\tev}\right)^2.
\label{higgsino}
\eeq
Using the 2-$\sigma$ range of dark matter density preferred after WMAP
data~\cite{wmap}
\beq
   0.094<\Omega_{DM}h^2< 0.129,
\label{wmap}
\eeq
we find that a higgsino dark matter should have a mass in the range
1.0 to 1.2 TeV.

Next we consider a $W$-ino lightest neutralino.
The W-ino has also a neutral component that could play the r\^ole of the
dark matter. It is not usually considered a standard candidate, because
in ordinary low-energy supersymmetry with gaugino unification condition,
it can never be the lightest state. However, the W-ino can become the
LSP in anomaly mediation~\cite{gher}. 
Both tree-level and one-loop effects
contribute to make the neutral state belonging to the $SU(2)$ triplet
lighter than the charged one~\cite{gher}.
In the limit of pure W-ino, the relic abundance in Split Supersymmetry is 
\beq
\Omega_{\tilde W}h^2=0.02~\left( \frac{M_2}{\tev}\right)^2.
\eeq
Using \eq{wmap}, we find that the mass range of a W-ino dark matter is
2.0 to 2.5 TeV. 

\begin{figure}
\begin{center}
\epsfig{file=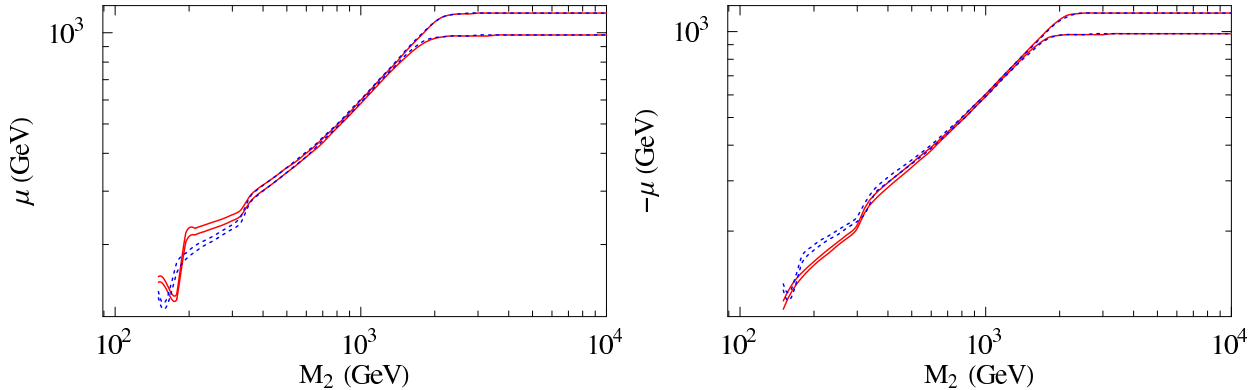,width=\textwidth}
\end{center}
\caption{The bands give the values of $\mu$, as functions of $M_2$,
consistent with the dark matter constraint $0.094<\Omega_{DM}h^2< 0.129$,
with gaugino mass unification
The solid lines correspond to $\tan\beta =4$ and the dashed lines to
$\tan\beta =20$. The left frame is for positive values of $\mu$ and the
right frame for negative values of $\mu$.}
\label{fig2}
\end{figure}

Let us now study the dependence of the neutralino relic abundance 
as we vary the parameters of Split Supersymmetry. The numerical
result of the correlation between $\mu$ and $M_2$ is shown in 
\fig{fig2}, assuming gaugino mass unification, $\tilde m =10^5~\gev$
and $\tan\beta =4$ and 20. The ratio $M_3/M_2$ is determined by the
equations discussed in sect.~\ref{secgmas}.

For small values of $M_2$, the lightest neutralino is mainly $B$-ino, but
as explained before, the higgsino component is non-negligible. When
$M_2$ is larger than the gauge-boson masses, \eq{bino} is a good approximation
and therefore the correlation between $M_2$ and $\mu$ is roughly
$\mu \propto M_2^{2/3}$, 
in agreement with the results 
shown in \fig{fig2}. As $M_2$ grows, the gap between $\mu$ and $M_1$
gets reduces until we have the transition to the higgsino region,
where the value of $\mu$ is uniquely determined by \eq{higgsino}.
Thus the behaviour of the curves in \fig{fig2} is well described by the
various limiting approximations.

\begin{figure}
\begin{center}
\epsfig{file=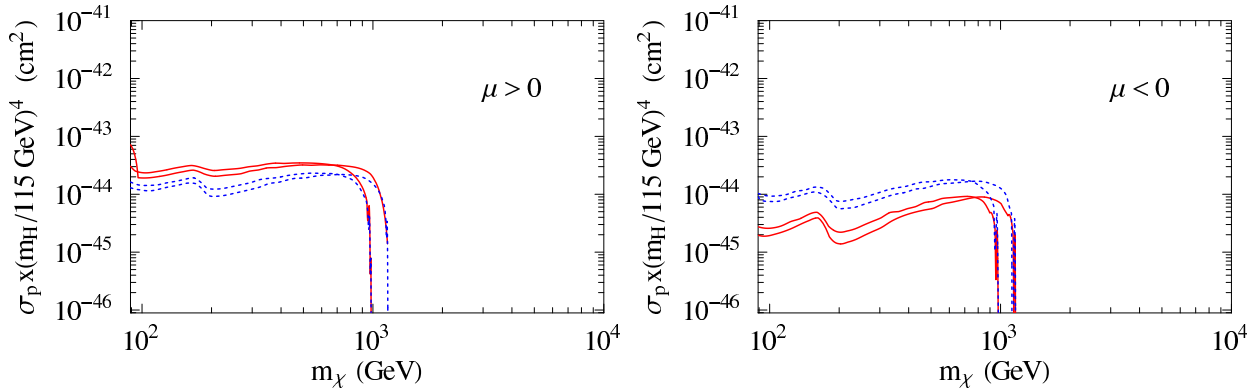,width=\textwidth}
\end{center}
\caption{The bands give neutralino-proton cross section relevant
for dark matter detection, as a function of the neutralino mass 
$m_\chi$. The cross section, which scales as $\sigma_p \propto m_H^{-4}$,
has been calculated for $m_H=115~\gev$.
The parameters are such that the neutralino relic abundance is 
consistent with the dark matter constraint $0.094<\Omega_{DM}h^2< 0.129$,
assuming gaugino mass unification.
The solid lines correspond to $\tan\beta =4$ and the dashed lines to
$\tan\beta =20$. 
The left frame is for positive values of $\mu$ and the
right frame for negative values of $\mu$.}
\label{fig3}
\end{figure}

Dark matter particles are actively searched for in underground experiments
through their scatterings with nuclei. In the case of
Split-Supersymmetry,
since squarks and sleptons are very heavy,
and the $Z$--$\chi^0$--$\chi^0$ coupling gives only a spin-dependent neutralino
interaction with nuclei, the only contribution to the spin-independent
cross section comes from the exchange of the Higgs boson~\cite{barg}.
To compare different experiments, it is customary to consider the neutralino 
scattering cross-section off a proton, which is given by
\beq
\sigma_p =
\left( N_{11}\tan\theta_W -N_{12}\right)^2
\left( N_{13}\cos\beta -N_{14}\sin\beta \right)^2
 \left( \frac{115~\gev}
{m_H}\right)^4 ~4\times 10^{-43}~{\rm cm}^2.
\label{crossh}
\eeq
Here $m_H$ is the Higgs mass and $N_{1,i}$ are the gaugino and higgsino
components of the lightest neutralino, in standard notations. As shown
by \eq{crossh}, the $H$--$\chi^0$--$\chi^0$ coupling vanishes when the
neutralino is a pure gaugino ($N_{13}=N_{14}=0$)
or pure higgsino ($N_{13}=N_{14}=0$). Interestingly, Split
Supersymmetry predicts that $\chi^0$ is a mixed state, as long as $M_2$
is not too large (see \fig{fig2}). However, the scattering rate is rather
small and it drops with the Higgs mass as $M_H^{-4}$. In \fig{fig3}, we
show the spin-independent neutralino-proton cross section for the same
parameter choice as in \fig{fig2}. We have normalized the cross
section with a fixed value of $m_H$, to allow a simple scaling of the
results. 
Notice how the detection rate is negligible when $\chi^0$
becomes higgsino ($M_2\gsim \tev$). Nevertheless, in the case of a mixed
state and in a large 
range of the allowed Higgs mass, the signal
is within the reach of future experiments, which will have
a sensitivity up to $10^{-44}$--$10^{-45}$~cm$^2$ for $m_\chi =1~\tev$. 

\subsection{Collider Signals}

Our dark matter analysis has identified three preferred regions of Split 
Supersymmetry. 

The first region is, for gaugino mass unification, when $M_2$ is 
below about 1--2 TeV, and the parameter 
$\mu$ is determined to be comparable (but slightly larger) than $M_1$. 
As $M_2$ grows,
the degeneracy between $\mu$ and $M_1$ becomes more pronounced.
The lightest neutralino is always a mixed state. 
In this region, the gaugino mass parameters can be as low as their
present experimental lower limit, giving particularly interesting prospects
for future searches.

The second case is when $\mu \simeq$~1.0--1.2~TeV and the gaugino masses
are (arbitrarily) larger than $\mu$. This is, for instance, the situation
when the $\mu$ parameter is radiatively generated, as considered in 
sect.~\ref{secgmas}. New particles can be considerably heavy. 

Finally, there is the possibility that $M_2\simeq$~2.0--2.5~TeV and the
other parameters are (arbitrarily) larger: this is the case with anomaly
mediation. 
It is not a very pleasent scenario for collider experiments, since
all new particles are very heavy. We want to stress however that, once 
we abandon the hierarchy problem, a lighter spectrum is not more natural
than a heavy one. All solutions to the dark matter requirements are
equally acceptable.

At the LHC, experiments can search for displaced vertices (if $\tilde m$
is low and the gluino decays inside the detector) or for stable
coloured particles (for larger $\tilde m$)~\cite{baer}. We expect that gluinos
should be visible for masses up to about 2.5~TeV. 
This limits the reach on the parameter
$M_2$ and certainly not all the region relevant
for dark matter will be covered. 
Therefore, the search for neutralinos and charginos
can be quite important.
In Split Supersymmetry, neutralinos and charginos are produced at the LHC
only through gauge-boson interaction, since there are no decay chains
originated by gluino and squark decays. The trilepton channel is also
not feasible, because of the small leptonic branching ratios of gauginos,
in the limit of heavy sleptons. 

There are two characteristics of Split Supersymmetry, which determine the
collider phenomenology of neutralinos and charginos. {\it (i)} In the
effective theory, the $B$-ino interacts only with Higgs and higgsinos,
see \eq{lagr}. Therefore, when $M_1$ is smaller than $M_2$ and $\mu$, 
we expect a significant
probability of finding Higgs bosons in the final states of  
the decay chains of the
heavy neutralinos and charginos. A complication exists when $M_1$ and 
$\mu$ become nearly mass degenerate. {\it (ii)} The phenomenology of
Split Supersymmetry is different than the one of 
ordinary low-energy supersymmetry
with heavy squarks, because of the value of the $\mu$ parameter. In the
ordinary case, radiative electroweak breaking leads to a large $\mu$ and
almost pure gauginos are the lightest states. In Split Supersymmetry $\mu$
is determined by dark matter considerations and higgsinos play an 
important r\^ole.

Identification of neutralinos and charginos at the LHC requires the
analysis of signals with multi-Higgs and missing energy in the final state.
Once the Higgs is discovered in the conventional channel and its mass 
determined, one can isolate multi-$\bar b$-$b$ pairs with the correct invariant
mass from the background. 

Linear colliders are most appropriate to extend the search for
neutralinos and charginos. In particular, a multi-TeV collider like
CLIC could entirely cover the cases of mixed-state or higgsino dark matter.
The $W$-ino case requires center-of-mass energies up to 4--5~TeV.
Moreover, a linear collider can determine the mass parameters and the
gaugino couplings $\tilde g$, testing the relations of Split Supersymmetry.

{\bf Acknowledgments} We wish to thank T.~Jones, M.~Mangano,
T.~Plehn, and R.~Rattazzi
for useful discussions.
  
{\bf Note added} After completion of this paper, the paper in 
ref.~\cite{stronz} appeared, giving the renormalization-group equations
for Split Supersymmetry at one-loop order. The equations for the gaugino 
coupling $\gtilu$, for the weak-gaugino mass, and for the Higgs coupling do
not agree with the results given in our appendix. Our results correctly
match the supersymmetric case.

\section*{Appendix}

In this appendix we give the renormalization-group equations for Split
Supersymmetry. They have been derived from the general expressions given in 
ref.~\cite{mac}. We have checked that when the contribution from heavy scalars
is included, our equations reproduce the supersymmetric result, given in
ref.~\cite{mart}.

The 2-loop renormalization-group equation for the gauge couplings is
\bea
(4\pi)^2\frac{d}{dt}~ g_i=g_i^3b_i &+&\frac{g_i^3}{(4\pi)^2}
\left[ \sum_{j=1}^3B_{ij}g_j^2-\sum_{\alpha=u,d,e}d_i^\alpha
{\rm Tr}\left( h^{\alpha \dagger}h^{\alpha}\right)\right. \nonumber \\
&-&\left. d_W\left( 
\gtiluq +\gtildq \right)
-d_B\left( \gtilupq +\gtildpq \right)\right] ,
\label{gauger}
\eea
where $t=\ln \bar \mu$ and $\bar \mu$ is the renormalization scale.
We use the convention $g_1^2=(5/3)g^{\prime 2}$.
Equation~(\ref{gauger}) is scheme-independent
up to the two-loop order.
 
In the effective theory below $\tilde{m}$, the $\beta$-function coefficients
are 
\bea
&&b=\left(\frac{9}{2},-\frac{7}{6},-5\right) ~~~~~B=\pmatrix{\frac{104}{25}&
\frac{18}{5}&\frac{44}{5}\cr \frac{6}{5} & \frac{106}{3}&12 \cr
\frac{11}{10}&\frac{9}{2}&22} \\
&&d^u=\left(\frac{17}{10},\frac{3}{2},2\right) ~~~~
d^d=\left(\frac{1}{2},\frac{3}{2},2\right) ~~~~
d^e=\left(\frac{3}{2},\frac{1}{2},0\right) \\
&&d^W=\left(\frac{9}{20},\frac{11}{4},0\right) ~~~~
d^B=\left(\frac{3}{20},\frac{1}{4},0\right) .
\eea

Above the scale $\tilde{m}$, we recover the supersymmetric result (identifying
$h^\alpha$ with $\lambda^{\alpha *}$)
\bea
&&b=\left(\frac{33}{5},1,-3\right) ~~~~~B=\pmatrix{\frac{199}{25}&
\frac{27}{5}&\frac{88}{5}\cr \frac{9}{5} & 25&24 \cr
\frac{11}{5}&9&14} \\
&&d^u=\left(\frac{26}{5},6,4\right) ~~~~
d^d=\left(\frac{14}{5},6,4\right) ~~~~
d^e=\left(\frac{18}{5},2,0\right) ~~~~d^W=
d^B=0 .
\eea

Below the mass of the lightest neutralino, the theory coincides with the
SM with one Higgs doublet and
\bea
&&b=\left(\frac{41}{10},-\frac{19}{6},-7\right) ~~~~~
B=\pmatrix{\frac{199}{50}&
\frac{27}{10}&\frac{44}{5}\cr \frac{9}{10} & \frac{35}{6}&12 \cr
\frac{11}{10}&\frac{9}{2}&-26} \\
&&d^u=\left(\frac{17}{10},\frac{3}{2},2\right) ~~~~
d^d=\left(\frac{1}{2},\frac{3}{2},2\right) ~~~~
d^e=\left(\frac{3}{2},\frac{1}{2},0\right) ~~~~d^W=
d^B=0 .
\eea

Since Yukawa and gaugino couplings appear in \eq{gauger} only in the 
two-loop part, we need their evolution up to the one-loop order.
For the Yukawa couplings, we find
\bea
(4\pi)^2\frac{d}{dt}~h^u&=&h^u\left( -3\sum_{i=1}^3c_i^ug_i^2
+\frac{3}{2}
h^{u \dagger}h^{u}
-\frac{3}{2}
h^{d \dagger}h^{d}
+T\right)\\
(4\pi)^2\frac{d}{dt}~h^d&=&h^d\left( -3\sum_{i=1}^3c_i^dg_i^2
-\frac{3}{2}
h^{u \dagger}h^{u}
+\frac{3}{2}
h^{d \dagger}h^{d}
+T\right)\\
(4\pi)^2\frac{d}{dt}~h^e&=&h^e\left( -3\sum_{i=1}^3c_i^eg_i^2
+\frac{3}{2}
h^{e \dagger}h^{e}
+T\right) .
\label{yukk}
\eea
Below the scale $\tilde{m}$, we find
\beq
T={\rm Tr} \left( 3h^{u \dagger}h^{u}+3 h^{d \dagger}h^{d}+
h^{e \dagger}h^{e}\right) +\frac{3}{2}\left( \gtiluq +\gtildq \right)
+\frac{1}{2}\left( \gtilupq +\gtildpq \right)
\eeq
\beq
c^u=\left( \frac{17}{60}, \frac{3}{4}, \frac{8}{3}\right) ~~~~
c^d=\left( \frac{1}{12}, \frac{3}{4}, \frac{8}{3}\right) ~~~~
c^e=\left( \frac{3}{4}, \frac{3}{4}, 0\right) .
\eeq
This result is valid also for the SM, if we take
$T={\rm Tr} ( 3h^{u \dagger}h^{u}+3 h^{d \dagger}h^{d}+
h^{e \dagger}h^{e})$.

Above the scale $\tilde{m}$, the renormalization-group equations of the
Yukawa couplings are
\bea
(4\pi)^2\frac{d}{dt}~\lambda^u&=&\lambda^u
\left[ -2\sum_{i=1}^3c_i^ug_i^2
+3\lambda^{u \dagger}\lambda^{u}+\lambda^{d \dagger}
\lambda^{d}+3{\rm Tr}(\lambda^{u \dagger}\lambda^{u}) \right]\\
(4\pi)^2\frac{d}{dt}~\lambda^d&=&\lambda^d
\left[ -2\sum_{i=1}^3c_i^dg_i^2
+\lambda^{u \dagger}\lambda^{u}+3\lambda^{d \dagger}
\lambda^{d}+{\rm Tr}(3\lambda^{d \dagger}\lambda^{d}
+\lambda^{e \dagger}\lambda^{e}) \right]\\
(4\pi)^2\frac{d}{dt}~\lambda^e&=&\lambda^e
\left[ -2\sum_{i=1}^3c_i^eg_i^2
+3\lambda^{e \dagger}\lambda^{e}+{\rm Tr}(3\lambda^{d \dagger}\lambda^{d}
+\lambda^{e \dagger}\lambda^{e}) \right] .
\eea
\beq
c^u=\left( \frac{13}{30}, \frac{3}{2}, \frac{8}{3}\right) ~~~~
c^d=\left( \frac{7}{30}, \frac{3}{2}, \frac{8}{3}\right) ~~~~
c^e=\left( \frac{9}{10}, \frac{3}{2}, 0\right) .
\eeq

The renormalization-group equations for the gaugino couplings defined
in \eq{lagr} are
\bea
(4\pi)^2\frac{d}{dt}~\gtilu &=&-3\gtilu \sum_{i=1}^3 C_i g_i^2
+\frac{5}{4}\gtiluc -\frac{1}{2} \gtilu \gtildq 
+\frac{1}{4} \gtilu \gtilupq
+\gtild \gtildp \gtilup +\gtilu T \\
(4\pi)^2\frac{d}{dt}~\gtilup &=&-3\gtilup \sum_{i=1}^3 C_i^{\prime} g_i^2
+\frac{3}{4}\gtilupc +\frac{3}{2} \gtilup \gtildpq 
+\frac{3}{4} \gtilup \gtiluq
+3\gtildp \gtild \gtilu +\gtilup T \\
(4\pi)^2\frac{d}{dt}~\gtild &=&-3\gtild \sum_{i=1}^3 C_i g_i^2
+\frac{5}{4}\gtildc -\frac{1}{2} \gtild \gtiluq 
+\frac{1}{4} \gtild \gtildpq
+\gtilu \gtilup \gtildp +\gtild T \\
(4\pi)^2\frac{d}{dt}~\gtildp &=&-3\gtildp \sum_{i=1}^3 C_i^{\prime} g_i^2
+\frac{3}{4}\gtildpc +\frac{3}{2} \gtildp \gtilupq 
+\frac{3}{4} \gtildp \gtildq
+3\gtilup \gtilu \gtild +\gtildp T ,
\eea
\beq
C=\left( \frac{3}{20},\frac{11}{4},0\right) ~~~~
C^\prime =\left( \frac{3}{20},\frac{3}{4},0\right) .
\eeq

The renormalization-group equations for the gaugino masses and the
$\mu$ parameter below the scale
$\tilde{m}$ are
\bea
(4\pi)^2\frac{d}{dt}~M_3 &=& -18g_3^2M_3 \left( 1+\frac{c_{\tilde g}
g_3^2}{(4\pi)^2}\right)
\\
(4\pi)^2\frac{d}{dt}~M_2 &=& \left( -12g_2^2+
\gtiluq + \gtildq \right) M_2 +4\gtilu \gtild \mu
\\
(4\pi)^2\frac{d}{dt}~M_1 &=& \left( 
\gtilupq + \gtildpq \right) M_1 +4\gtilup \gtildp \mu
\\ 
(4\pi)^2\frac{d}{dt}~\mu &=& \frac 14\left[
-18\left( \frac{g_1^2}{5}+g_2^2\right) 
+3\left( \gtiluq +\gtildq \right) 
+\gtilupq  +\gtildpq \right]
\mu + 3\gtilu \gtild M_2 +\gtilup \gtildp M_1.
\eea
For $M_3$ we have included also the next-to-leading order correction and
we find $c_{\tilde g}=38/3$ in $\overline{\rm MS}$ and $c_{\tilde g}=10$
in $\overline{\rm DR}$. The relation between the gluino running and pole mass
is
\beq
M_{\tilde{g}}^{\rm pole} = M_3(\bar \mu )\left[ 1+\frac{g_3^2}{(4\pi )^2}
\left( C_{\tilde g} +9\ln \frac{{\bar \mu}^2}{M_3^2}\right) \right] ,
\eeq
with $C_{\tilde g}=12$ in $\overline{\rm MS}$ and $c_{\tilde g}=15$
in $\overline{\rm DR}$.

Above the scale $\tilde{m}$, the equations for the gaugino masses and
$\mu$ are given by \bea
(4\pi)^2\frac{d}{dt}~M_i &=& 2b_i g_i^2 M_i \\
(4\pi)^2\frac{d}{dt}~\mu &=& \left[ -3g_2^2-\frac 35 g_1^2 + {\rm
    Tr}(3\lambda^{u \dagger}\lambda^{u}+ 3\lambda^{d
    \dagger}\lambda^{d} +\lambda^{e \dagger}\lambda^{e})\right] \mu .
\eea The two-loop expression of the gluino mass can be found in
ref.~\cite{mart}. In our numerical analysis we have set to zero the
trilinear $A$-terms.

Finally, the equation for the Higgs quartic coupling is
\bea
(4\pi)^2\frac{d}{dt}~\lambda  &=&12 \lambda^2 +\lambda \left[
-9 \left( \frac{g_1^2}{5}+g_2^2\right)+6\left( \gtiluq +\gtildq \right)
+2\left( \gtilupq +\gtildpq \right) \right. \nonumber \\
&&\left. +4 {\rm Tr}(3h^{u \dagger}h^{u}+
3h^{d \dagger}h^{d}
+h^{e \dagger}h^{e})\right] +\frac 92 \left( \frac{g_2^4}{2}+
\frac{3g_1^4}{50}+\frac{g_1^2g_2^2}{5}
\right) \nonumber \\
&& -5\left( \gtiluqq +\gtildqq \right) -2 \gtiluq\gtildq
-\left( \gtilupq +\gtildpq \right)^2  -2
\left( \gtilu \gtilup + \gtild \gtildp \right)^2 \nonumber \\
&&-4 {\rm Tr}\left[3(h^{u \dagger}h^{u})^2+
3(h^{d \dagger}h^{d})^2
+(h^{e \dagger}h^{e})^2 \right] .
\eea



\begin{thebibliography}{20}

\bibitem{wein}
A.~Vilenkin,
Phys.\ Rev.\ Lett.\  {\bf 74} (1995) 846
[arXiv:gr-qc/9406010];
S.~Weinberg,
Phys.\ Rev.\ Lett.\  {\bf 59} (1987) 2607.

\bibitem{string}
R.~Bousso and J.~Polchinski,
JHEP {\bf 0006}, 006 (2000)
[arXiv:hep-th/0004134];
S.~Kachru, R.~Kallosh, A.~Linde and S.~P.~Trivedi,
Phys.\ Rev.\ D {\bf 68}, 046005 (2003)
[arXiv:hep-th/0301240];
L.~Susskind,
arXiv:hep-th/0302219;
F.~Denef and M.~R.~Douglas,
arXiv:hep-th/0404116.

\bibitem{savas}
N.~Arkani-Hamed and S.~Dimopoulos,
arXiv:hep-th/0405159.

\bibitem{alt}
G.~Altarelli and M.~W.~Grunewald,
arXiv:hep-ph/0404165.

\bibitem{lattice}
S.~Aoki {\it et al.}  [JLQCD Collaboration],
Phys.\ Rev.\ D {\bf 62}, 014506 (2000)
[arXiv:hep-lat/9911026].

\bibitem{kam}
Y.~Suzuki {\it et al.}  [TITAND Working Group Collaboration],
arXiv:hep-ex/0110005.

\bibitem{wagner}
D.~Choudhury, T.~M.~P.~Tait and C.~E.~M.~Wagner,
Phys.\ Rev.\ D {\bf 65}, 053002 (2002)
[arXiv:hep-ph/0109097];
D.~E.~Morrissey and C.~E.~M.~Wagner,
Phys.\ Rev.\ D {\bf 69}, 053001 (2004)
[arXiv:hep-ph/0308001].

\bibitem{altri}
L.~E.~Ibanez,
Phys.\ Lett.\ B {\bf 126} (1983) 196
[Erratum-ibid.\ B {\bf 130} (1983) 463];
J.~E.~Bjorkman and D.~R.~T.~Jones,
Nucl.\ Phys.\ B {\bf 259} (1985) 533;
T.~G.~Rizzo,
Phys.\ Rev.\ D {\bf 45}, 3903 (1992).


\bibitem{giu}
G.~F.~Giudice and A.~Masiero,
Phys.\ Lett.\ B {\bf 206}, 480 (1988).

\bibitem{baer}
H.~Baer, K.~m.~Cheung and J.~F.~Gunion,
Phys.\ Rev.\ D {\bf 59}, 075002 (1999)
[arXiv:hep-ph/9806361].

\bibitem{lim1}
P.~F.~Smith, J.~R.~J.~Bennett, G.~J.~Homer, J.~D.~Lewin, H.~E.~Walford and W.~A.~Smith,
Nucl.\ Phys.\ B {\bf 206}, 333 (1982).

\bibitem{lim2}
T.~K.~Hemmick {\it et al.},
Phys.\ Rev.\ D {\bf 41}, 2074 (1990).

\bibitem{sogl}
L.~J.~Hall,
Nucl.\ Phys.\ B {\bf 178}, 75 (1981).

\bibitem{ratt}
L.~J.~Hall, R.~Rattazzi and U.~Sarid,
Phys.\ Rev.\ D {\bf 50}, 7048 (1994)
[arXiv:hep-ph/9306309].

\bibitem{neutr}
F.~Vissani and A.~Y.~Smirnov,
Phys.\ Lett.\ B {\bf 341} (1994) 173
[arXiv:hep-ph/9405399];
A.~Brignole, H.~Murayama and R.~Rattazzi,
Phys.\ Lett.\ B {\bf 335} (1994) 345
[arXiv:hep-ph/9406397].

\bibitem{obl}
H.~C.~Cheng, J.~L.~Feng and N.~Polonsky,
Phys.\ Rev.\ D {\bf 56}, 6875 (1997)
[arXiv:hep-ph/9706438];
E.~Katz, L.~Randall and S.~f.~Su,
Nucl.\ Phys.\ B {\bf 536} (1998) 3
[arXiv:hep-ph/9801416];
S.~Kiyoura, M.~M.~Nojiri, D.~M.~Pierce and Y.~Yamada,
Phys.\ Rev.\ D {\bf 58} (1998) 075002
[arXiv:hep-ph/9803210].

\bibitem{anom}
L.~Randall and R.~Sundrum,
Nucl.\ Phys.\ B {\bf 557}, 79 (1999)
[arXiv:hep-th/9810155];
G.~F.~Giudice, M.~A.~Luty, H.~Murayama and R.~Rattazzi,
JHEP {\bf 9812}, 027 (1998)
[arXiv:hep-ph/9810442].

\bibitem{hig}
R.~Barate {\it et al.},
Phys.\ Lett.\ B {\bf 565} (2003) 61
[arXiv:hep-ex/0306033].

\bibitem{coan}
J.~Edsjo and P.~Gondolo,
Phys.\ Rev.\ D {\bf 56} (1997) 1879
[arXiv:hep-ph/9704361].

\bibitem{ds}
P.~Gondolo, J.~Edsjo, P.~Ullio, L.~Bergstrom, M.~Schelke and E.~A.~Baltz,
arXiv:astro-ph/0211238.

\bibitem{wmap}
C.~L.~Bennett {\it et al.},
Astrophys.\ J.\ Suppl.\  {\bf 148} (2003) 1
[arXiv:astro-ph/0302207];
D.~N.~Spergel {\it et al.},
Astrophys.\ J.\ Suppl.\  {\bf 148} (2003) 175
[arXiv:astro-ph/0302209].

\bibitem{gher}
T.~Gherghetta, G.~F.~Giudice and J.~D.~Wells,
Nucl.\ Phys.\ B {\bf 559}, 27 (1999)
[arXiv:hep-ph/9904378].

\bibitem{barg}
R.~Barbieri, M.~Frigeni and G.~F.~Giudice,
Nucl.\ Phys.\ B {\bf 313}, 725 (1989).

\bibitem{mac}
M.~E.~Machacek and M.~T.~Vaughn,
Nucl.\ Phys.\ B {\bf 222}, 83 (1983);
Nucl.\ Phys.\ B {\bf 236}, 221 (1984);
Nucl.\ Phys.\ B {\bf 249}, 70 (1985).

\bibitem{mart}
S.~P.~Martin and M.~T.~Vaughn,
Phys.\ Rev.\ D {\bf 50}, 2282 (1994)
[arXiv:hep-ph/9311340].

\bibitem{stronz}
A.~Arvanitaki, C.~Davis, P.~W.~Graham and J.~G.~Wacker,
arXiv:hep-ph/0406034.





\end{thebibliography}
\end{document}